\documentclass[journal]{IEEEtran}

\usepackage{algorithm}
\usepackage{algpseudocode}
\usepackage{amsmath}
\usepackage{subfigure}

\usepackage{amssymb}
\usepackage{color}
\usepackage{graphicx}
\usepackage{cite}
\usepackage{multirow}
\usepackage{theorem}
\usepackage{url}
\usepackage{epstopdf}
\usepackage{etoolbox}
\usepackage{array}
\usepackage{lipsum}
\usepackage{footnote}
\usepackage[flushleft]{threeparttable}
\usepackage{booktabs}
\usepackage{cleveref}
\usepackage{mathtools}

\usepackage{enumitem}
\usepackage{soul}

\newcommand{\ignore}[1]{}

\newcommand{\blueHL}[1]{{\textcolor{blue}{#1}}}

\newcommand{\RN}[1]{%
  \textup{\uppercase\expandafter{\romannumeral#1}}%
}

\ifCLASSOPTIONcaptionsoff
  \usepackage[nomarkers]{endfloat}
 \let\MYoriglatexcaption\caption
 \renewcommand{\caption}[2][\relax]{\MYoriglatexcaption[#2]{#2}}
\fi

\usepackage{booktabs,fixltx2e}
\usepackage[table]{xcolor}
\usepackage{dblfloatfix}    

\usepackage{tikz}
\newcommand*\circled[1]{\tikz[baseline=(char.base)]{
            \node[shape=circle,draw,inner sep=2pt] (char) {#1};}}

\begin{document}

\IEEEoverridecommandlockouts
\IEEEpubid{\begin{minipage}[t]{\textwidth}\ \\[5pt]
        \centering\normalsize{\footnotesize{\copyright~2020 IEEE. Personal use of this material is permitted. Permission from IEEE must be obtained for all other uses, in any current or future media, including reprinting/republishing this material for advertising or promotional purposes, creating new collective works, for resale or redistribution to servers or lists, or reuse of any copyrighted component of this work in other works.}}
\end{minipage}}

\title{NeuPart: Using Analytical Models to Drive Energy-Efficient Partitioning of CNN Computations on Cloud-Connected Mobile Clients}
%
%
%

\author{Susmita Dey Manasi, Farhana Sharmin Snigdha, and Sachin S. Sapatnekar
\thanks{Accepted April 28, 2020 for publication in IEEE Transactions on Very Large Scale Integration (VLSI) Systems. This work was supported in part by the National Science Foundation (NSF) under Award CCF-1763761. (Corresponding author: Susmita Dey Manasi.)}
\thanks{The authors are with the Department of Electrical and Computer Engineering, University of Minnesota Twin Cities, Minneapolis, MN 55455 USA (e-mail: manas018@umn.edu; sharm304@umn.edu; sachin@umn.edu).}
\thanks{Digital Object Identifier 10.1109/TVLSI.2020.2995135}
}

\maketitle

\begin{abstract}
Data processing on convolutional neural networks (CNNs) places a heavy burden
on energy-constrained mobile platforms. This work optimizes energy on a mobile
client by partitioning CNN computations between {\em in situ} processing on the
client and offloaded computations in the cloud.  A new analytical CNN energy
model is formulated, capturing all major components of the {\em in situ}
computation, for ASIC-based deep learning accelerators.  The model is
benchmarked against measured silicon data. The analytical framework is used to
determine the optimal energy partition point between the client and the cloud
at runtime. On standard CNN topologies, partitioned computation is demonstrated
to provide significant energy savings on the client over fully cloud-based or
fully {\em in situ} computation. For example, at 80 Mbps effective bit rate and 0.78 W
transmission power, the optimal partition for AlexNet [SqueezeNet] saves up to
52.4\% [73.4\%] energy over a fully cloud-based computation, and 27.3\% [28.8\%]
energy over a fully {\em in situ} computation.
\end{abstract}


%
\IEEEpeerreviewmaketitle

\section{Introduction}
\label{sec:intro}

\subsection{Motivation}

\noindent
Machine learning using deep convolutional neural networks (CNNs) constitutes a
powerful approach that is capable of processing a wide range of visual
processing tasks with high accuracy. Due to the highly energy-intensive nature
of CNN computations, today's deep learning (DL) engines using CNNs are largely
based in the cloud~\cite{Jouppi2017, Lee2017}, where energy is less of an issue
than on battery-constrained mobile clients. Although a few simple emerging
applications such as facial recognition are performed {\em in situ} on 
mobile processors, today's dominant mode is to offload DL computations from the
mobile device to the cloud. The deployment of specialized hardware accelerators
for embedded DL to enable energy efficient execution of CNN tasks is the next
frontier.

\begin{figure}[!ht]
\vspace{-0.2cm}
\centering
\includegraphics[width=2.4in]{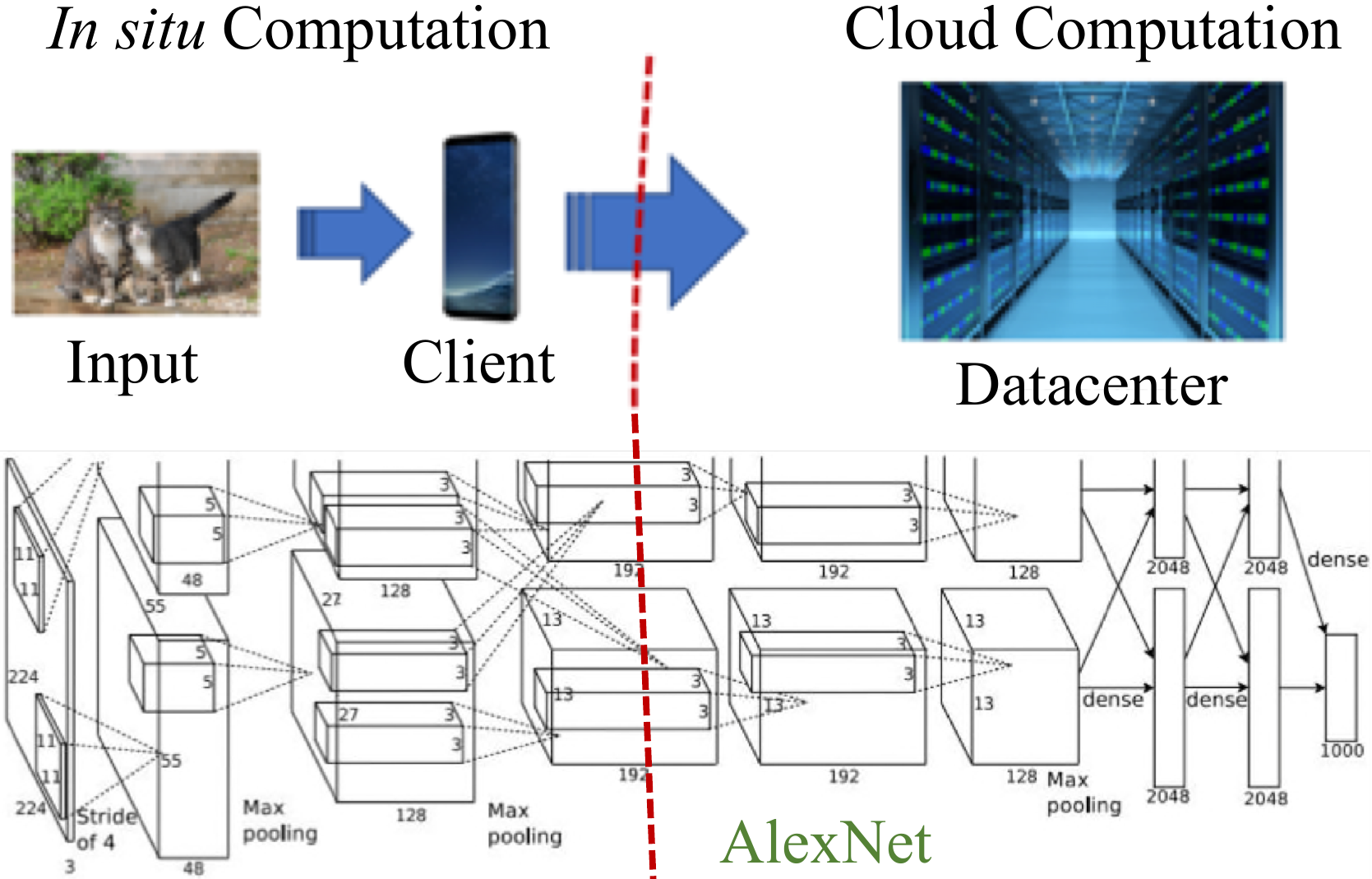}
\caption{An AlexNet computation, showing client/cloud partitioning.}
\label{fig:scheme}
\vspace{-0.6cm}
\end{figure}

Limited client battery life places stringent energy limitations on embedded DL
systems. This work focuses on a large class of inference engine applications
where battery life considerations are paramount over performance: e.g., for a
health worker in a remote area, who uses a mobile client to capture images
processed using DL to diagnose cancer~\cite{Esteva17}, a farmer who takes crop
photographs and uses DL to diagnose plant diseases~\cite{Mohanty16}, or an
unmanned aerial vehicle utilizing DL for monitoring populations of wild birds~\cite{Hong2019}.
In all these examples,
client energy is paramount for the operator in the field and
processing time is secondary, i.e., while arbitrarily long processing times are
unacceptable, somewhat slower processing times are acceptable for these
applications.  Moreover, for client-focused design, it is reasonable to assume
that the datacenter has plentiful power supply, and the focus is on minimizing
client energy rather than cloud energy.  While this paradigm may not apply to
all DL applications (e.g., our solution is not intended to be applied to
applications such as autonomous vehicles, where high-speed data processing is
important), the class of energy-critical client-side applications without
stringent latency requirements encompasses a large corpus of embedded DL tasks
that require energy optimization at the client end.

To optimize client energy, this work employs {\em computation partitioning}
between the client and the cloud.  Fig.~\ref{fig:scheme} shows an inference
engine computation on AlexNet~\cite{krizhevsky2012}, a representative CNN
topology, for recognition of an image from the camera of a mobile client.  If
the CNN computation is fully offloaded to the cloud, the image from the camera
is sent to the datacenter, incurring a communication overhead corresponding to
the number of data bits in the compressed image. Fully {\em in situ} CNN
computation on the mobile client involves no communication, but drains its
battery during the energy-intensive computation.  

\begin{figure}[!ht]
\vspace{-0.2cm}
\centering
\subfigure[Computation cost]{
\includegraphics[height=3.3cm]{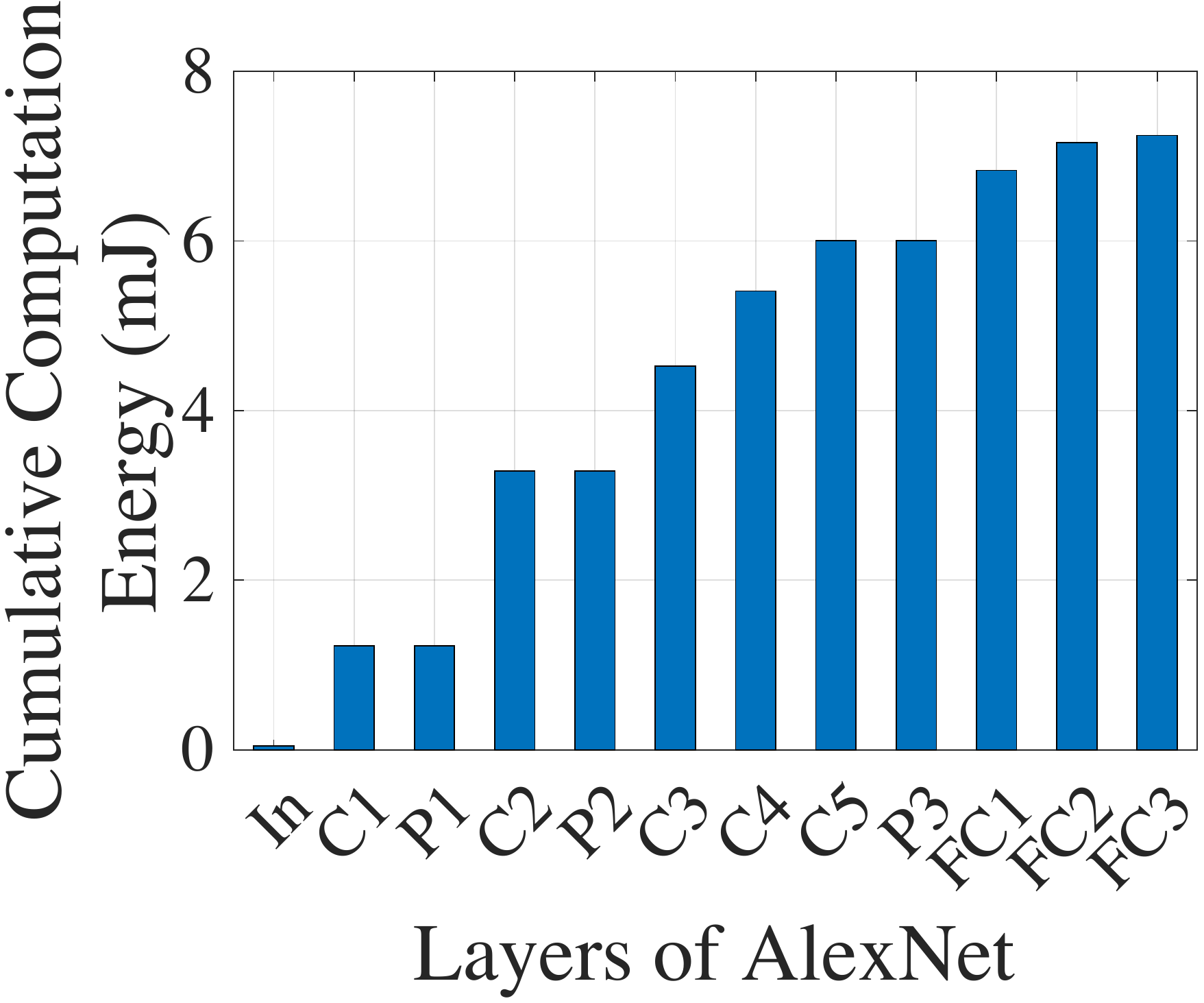}
}
\subfigure[Communication cost]{
\includegraphics[height=3.3cm]{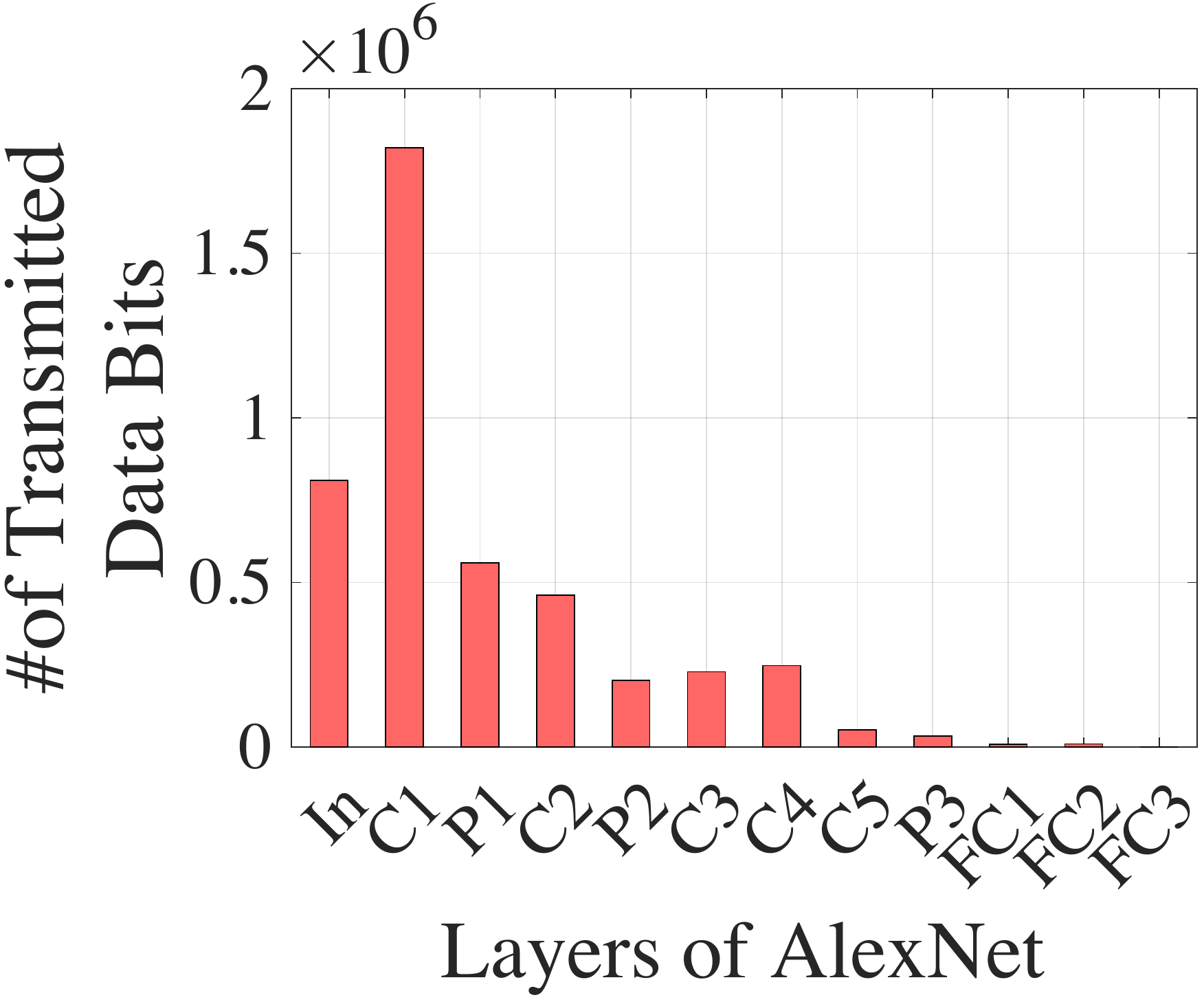}
}
\caption{(a) The cumulative computation energy up to each layer of AlexNet.
(b) Number of bits of compressed output data at each layer, to be
transmitted to the cloud.
}
\label{fig:compcomm}
\vspace{-0.4cm}
\end{figure}

Computation partitioning between the client and the cloud represents a middle
ground: the computation is partially processed {\em in situ}, up to a specific
CNN layer, on the client.  The data is then transferred to the cloud to
complete the computation, after which the inference results are sent back
to the client, as necessary. We propose NeuPart, a partitioner for DL tasks that minimizes
client energy.  NeuPart is based on an analytical modeling framework, and
analytically identifies, {\em at runtime}, the energy-optimal partition point
at which the partial computation on the client is sent to the cloud.

Fig.~\ref{fig:compcomm} concretely illustrates the tradeoff between
communication and computation for AlexNet. In (a), we show the cumulative
computation energy (obtained from our analytical CNN energy model presented in
Section~\ref{sec:AMsection}) from the input to a specific layer of the network.
In (b), we show the volume of compressed data that must be transmitted when the
partial {\em in situ} computation is transmitted to the cloud. The data
computed in internal layers of the CNN tend to have significant sparsity (over
80\%, as documented later in Fig.~\ref{fig:AvgStd}): NeuPart leverages this
sparsity to transmit only nonzero data at the point of partition, thus limiting
the communication cost.

The net energy cost for the client for a partition at the $L^{\rm th}$ layer
can be computed as:
\begin{eqnarray}
E_{Cost} = E_L + E_{Trans}
\label{eq:Ecost}
\end{eqnarray}
where $E_L$ is the processing energy on the client, up to the $L^{\rm th}$
CNN layer, and $E_{Trans}$ is the energy required to transmit this partially
computed data from the client to the cloud.  The inference result corresponds
to a trivial amount of data to return the identified class, and involves
negligible energy.  

NeuPart focuses on minimizing $E_{Cost}$ for the energy-constrained client.
For the data in Fig.~\ref{fig:compcomm}, $E_L$ increases monotonically as we
move deeper into the network, but $E_{Trans}$ can reduce greatly.  Thus,
the optimal partitioning point for $E_{Cost}$ here lies at an intermediate
layer $L$.

\subsection{Contributions of NeuPart}

\noindent
The specific contributions of NeuPart are twofold:
\begin{itemize}
\item	We develop a new analytical energy model (we name our model ``CNNergy" which is the core of NeuPart) to estimate the energy of executing CNN workload (i.e., $E_L$) on an ASIC-based deep learning (DL) accelerator. 
\item Unlike prior computation partitioning works~\cite{HLi2018,DNNSurgery2019,JALAD2018,IONN2018,
eshratifar2018,kang2017}, NeuPart addresses the client/cloud partitioning problem for specialized DL accelerators
that are more energy-efficient than CPU/GPU/FPGA-based platforms. 
\end{itemize}

\noindent
Today, neural networks are driving research in many fields and customized neural hardware is possibly the most active area in IC design research. While there exist many simulation platforms for performance analysis for general-purpose processors~\cite{Wattch2000}, memory systems~\cite{Wilton1996}, and NoCs~\cite{ORION2009}, there is no comparable performance simulator for ASIC implementations of CNNs.  Our model, CNNergy, is an attempt to fill that gap. CNNergy accounts for the complexities of scheduling computations
over multidimensional data. It captures key parameters of the hardware
and maps the hardware to perform computation on various CNN topologies. CNNergy is benchmarked on several
CNNs: AlexNet, SqueezeNet-v1.1~\cite{iandola2016}, VGG-16~\cite{Karen2014}, and GoogleNet-v1~\cite{Szegedy2015CVPR}. CNNergy is
far more detailed than prior energy models~\cite{Zhang18} and incorporates
implementation specific details of a DL accelerator, capturing all major
components of the {\em in situ} computation, including the cost of arithmetic
computations, memory and register access, control, and clocking. It is important to note that, for an ASIC-based neural accelerator platform, unlike a general-purpose processor, the computations are highly structured and the memory fetches are very predictable. Unlike CPUs, there are no conditionals or speculative fetches that can alter program flow significantly. Therefore, an analytical modeling framework (as validated in Section~\ref{sec:VAM}) is able to predict the CNN energy consumption closely for the custom ASIC platform.

CNNergy may potentially have utility beyond this
work, and has been open-sourced at\blueHL{~\url{https://github.com/manasiumn37/CNNergy}}. 
For example, it provides a breakdown of the total energy into specific
components, such as data access energy from different memory levels of a DL
accelerator, data access energy associated with each CNN data type from each
level of memory, MAC computation energy. CNNergy can also be used to explore
design phase tradeoffs such as analyzing the impact of changing on-chip memory
size on the total execution energy of a CNN. We believe that our developed simulator (CNNergy) will be useful to the practitioners who need an energy model for CNNs to evaluate various design choices. The application of data partitioning between client and cloud shows a way to apply our energy model to a practical scenario.

The paper is organized as follows. Section~\ref{sec:RelWork} discusses prior
approaches to computational partitioning and highlights the differences of NeuPart as compared to the prior works. In
Section~\ref{sec:bcgrnd}, fundamental background on CNN computation is provided, and the general framework of CNNergy for CNN energy estimation on custom ASIC-based DL accelerators is outlined. Next, Sections~\ref{sec:AMsection} presents
the detailed modeling of CNNergy and is followed by Section~\ref{sec:VAM}, which validates the model in several ways, including against silicon data. Section~\ref{sec:TrnDelay} presents the models for the estimation of transmission energy as well as inference delay.
A method for performing the NeuPart client/cloud partitioning at runtime is discussed in
Section~\ref{sec:RunP}. Finally, in Section~\ref{sec:res}, the evaluations of the client/cloud partitioning using NeuPart is presented under various communication environments for widely used CNN topologies. The paper concludes in Section~\ref{sec:Conclu}.

\section{Related Work}
\label{sec:RelWork}

\noindent
Computational partitioning has previously been used in the general context of
distributed processing~\cite{Kumar2010}.
A few prior works~\cite{HLi2018,DNNSurgery2019,JALAD2018,IONN2018,
eshratifar2018,kang2017} have utilized computation partitioning in the context
of mobile DL. In~\cite{HLi2018}, tasks are offloaded to a server from nearby
IoT devices for best server utilization, but no attempt is made to minimize
edge device energy. In~\cite{DNNSurgery2019} and~\cite{JALAD2018}, partitioning is used to optimize overall delay or throughput for delay critical applications of deep neural network (DNN) (e.g., self-driving cars~\cite{DNNSurgery2019}). Another work,~\cite{IONN2018}, uses partitioning between the client and local server (in contrast to centralized cloud) where along with the inference data, the client also needs to upload the partial DNN model (i.e., DNN weights) to the local server every time it makes an inference request. Therefore, the optimization goals and the target platforms of these works are very different from NeuPart.

The work in~\cite{eshratifar2018} uses limited
application-specific profiling data to schedule computation partitioning.
Another profiling-based scheme~\cite{kang2017} uses client-specific profiling
data to form a regression-based computation partitioning model for each device. 
A limitation of profiling-based approaches is that they require profiling data
for each mobile device or each DL application, which implies that a large
number of profiling measurements are required for real life deployment.
Moreover, profiling-based methods require the hardware to already be deployed
and cannot support design-phase optimizations.  Furthermore, all these prior
approaches use a CPU/GPU-based platform for the execution of DL workloads. 

In contrast with prior methods, NeuPart works with specialized DL accelerators,
which are orders of magnitude more energy-efficient as compared to the general-purpose machines~\cite{Jouppi2017,TChen2015,shidiannao2015}, for client/cloud
partitioning.  NeuPart specifically leverages the highly structured nature of
computations on CNN accelerators, and shows that an analytical model predicts
the client energy accurately (as demonstrated in Section~\ref{sec:VAM}). 
The analytical framework used in the NeuPart CNNergy incorporates
implementation-specific details that are not modeled in prior works. 
For example, the work in Neurosurgeon~\cite{kang2017} uses (a) uncompressed raw image to transmit at the
input, which is not typical: in a real system, images are compressed
before transmission to reduce the communication overhead; (b) unequal bit width
(i.e., 32-bit data for the intermediate layers while 8-bit data for the input layer; and (c) ignores any data sparsity
at the intermediate CNN layers.
Consequently, these cause the partitioning decision by Neurosurgeon to be either client-only or cloud-only in most cases.
In contrast, in addition to using a specialized DL accelerator, NeuPart fully leverages the inherent
computation-communication tradeoff of CNNs by exploiting their key properties and shows that (Fig.~\ref{fig:Quartile} in Section~\ref{sec:res}) there is a wide space where an intermediate partitioning point can offer significant energy savings as compared to the client-only or cloud-only approaches.

\section{Computations on CNN Hardware}
\label{sec:bcgrnd}

\subsection{Fundamentals of CNNs}
\label{ssec:cnn}

\begin{figure}[!ht]
\centering
\includegraphics[width=3.4in]{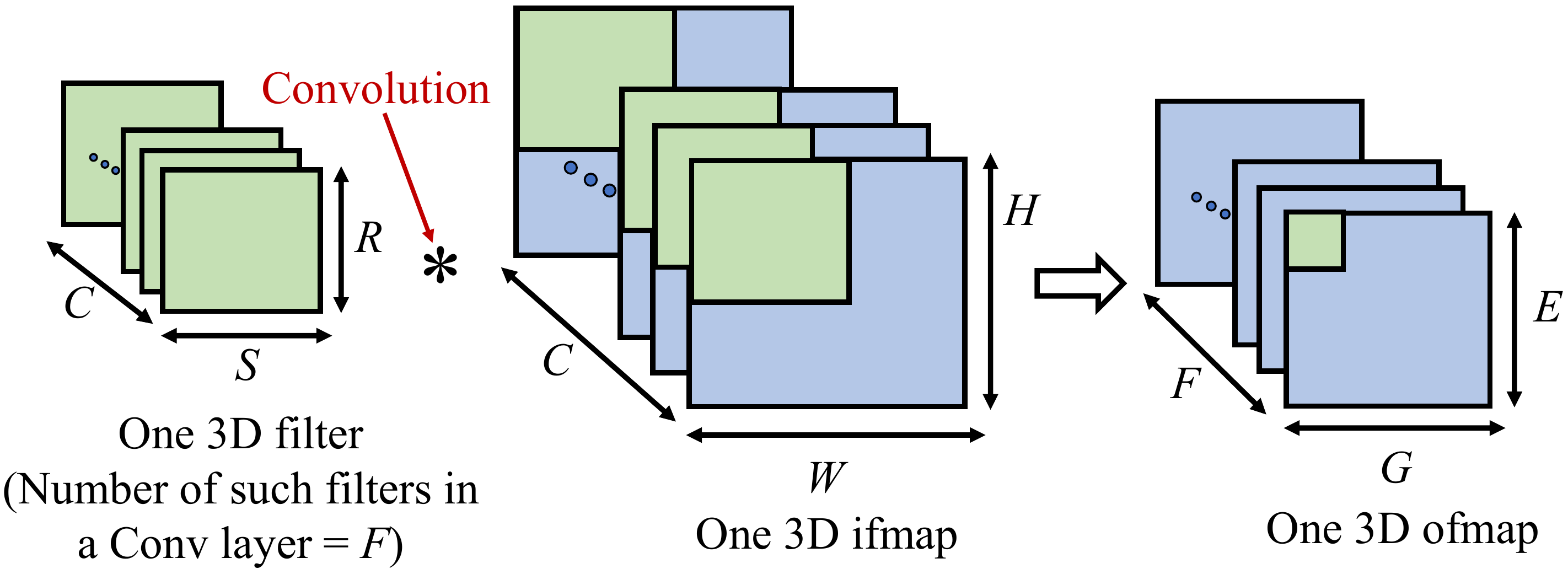}
\caption{Illustration of ifmap, filter, and ofmap in a Conv layer.}
\label{fig:ConvL}
\end{figure}

\begin{table}[!ht]
\vspace{-0.2cm}
\centering
\caption{Parameters for a CNN layer shape.}
\label{tbl:CNNParam}
\begin{tabular}{|c|l|}
\hline
Parameters & \multicolumn{1}{c|}{Description}                                 \\ \hline
$R/S$        	  & Height/width of a filter                                  \\ \hline
$H/W$        	  & Padded height/width of an ifmap                           \\ \hline
$E/G$        	  & Height/width of an ofmap                                  \\ \hline
$C$          	  & \#of channels in an ifmap and filter                      \\ \hline
$F$          	  & \begin{tabular}[c]{@{}l@{}}\#of 3D filters in a layer\\
                         \#of channels in an ofmap \end{tabular}              \\ \hline
$U$          	  & Convolution stride                                                                               												   \\ \hline
\end{tabular}
\vspace{-0.2cm}
\end{table}

\noindent 
The computational layers in CNNs can be categorized into three types:
convolution (Conv), fully connected (FC), and pooling (Pool).  The computation
in a CNN is typically dominated by the Conv layers. 
In each layer, the computation involves three types of data: 
\begin{itemize}
\item
{\bf ifmap}, the input feature map
\item
{\bf filter}, the filter weights, and
\item
{\bf psum}, the intermediate partial sums.
\end{itemize}

Table~\ref{tbl:CNNParam} summarizes the parameters associated with a
convolution layer. As shown in Fig.~\ref{fig:ConvL}, for a {\bf Conv layer},
filter and ifmap are both 3D data types consisting of multiple 2D planes
(channels).  Both the ifmap and filter have the same number of channels, $C$, while
$H \gg R$ and $W \gg S$.

During the convolution, an element-wise multiplication between the filter and
the green 3D region of the ifmap in Fig.~\ref{fig:ConvL} is followed by the
accumulation of all products (i.e., psums), and results in one element shown by
the green box in the output feature map (ofmap). Each channel ($R \times S
\times 1$) of the filter slides through its corresponding channel ($W \times H
\times 1$) of the ifmap with a stride ($U$), repeating similar
multiply-accumulate (MAC) operations to produce a full 2D plane ($E \times G
\times 1$) of the ofmap. A nonlinear activation function (e.g., a rectified
linear unit, ReLU) is applied after each layer, introducing sparsity (i.e.,
zeros) at the intermediate layers, which can be leveraged to reduce computation.

The above operation is repeated for all $F$ filters to produce $F$ 2D
planes for the ofmap, i.e., the number of channels in the ofmap equals the number of
3D filters in that layer.  Due to the nature of the convolution operation,
there is ample opportunity for data reuse in a Conv layer.  {\bf FC layers} are
similar to Conv layers but are smaller in size, and produce a 1D ofmap.  
Computations in the {\bf Pool layers} serve to reduce dimensionality of the ofmaps
produced from the Conv layers by storing the maximum/average value over a
window of the ofmap.

\subsection{Executing CNN Computations on a Custom ASIC}
\label{sec:cnnA}

\subsubsection{Architectural Features of CNN Hardware Accelerators}
\label{sec:arch_features}

\begin{figure}[!ht]
\centering
\includegraphics[width=3.2in]{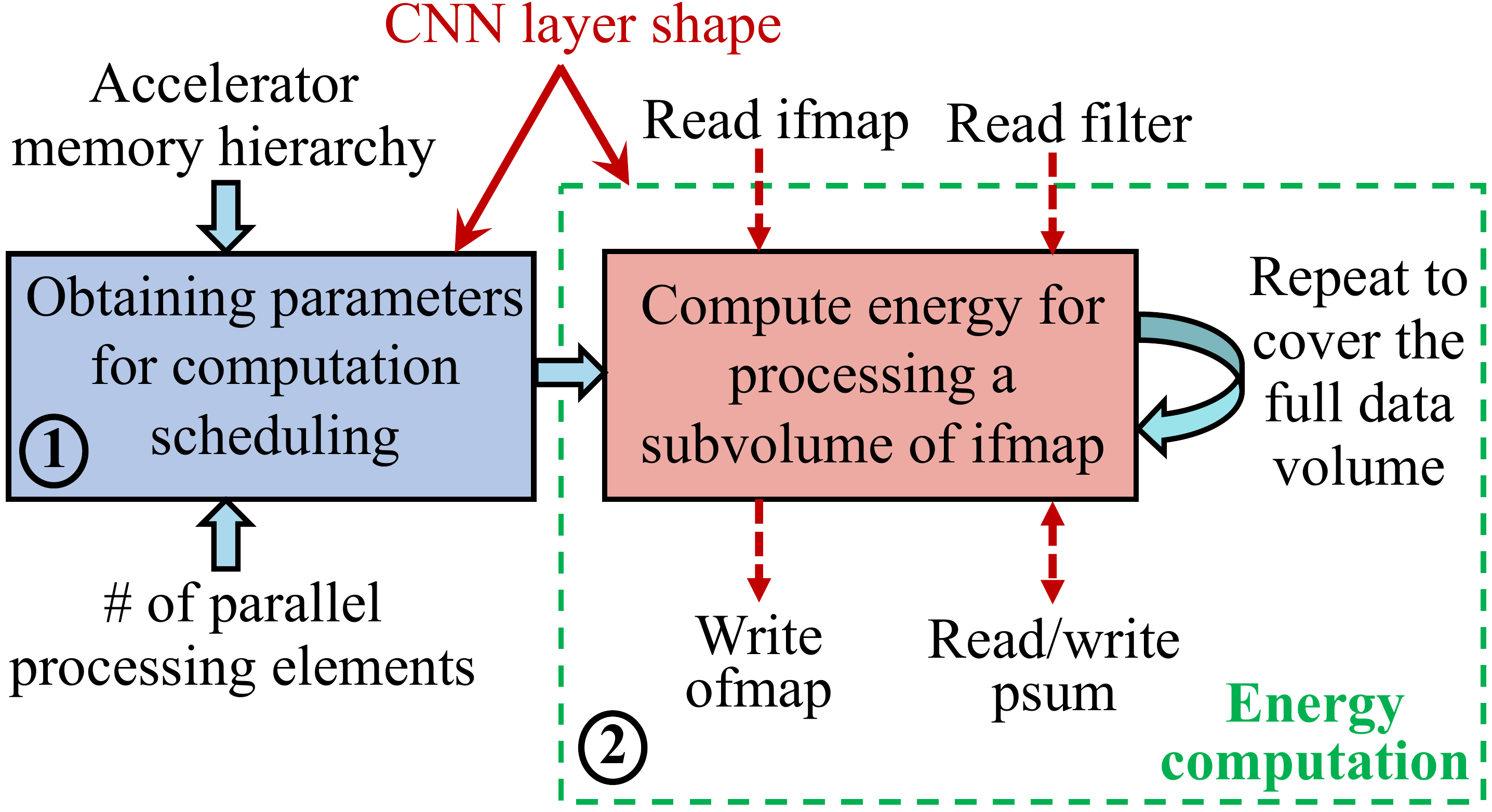}
\caption{General framework of the analytical CNN energy model (CNNergy).}
\label{fig:GenFrm}
\end{figure}

Since the inference task of a CNN comprises a very structured and fixed set
of computations (i.e., MAC, nonlinear activation, pooling,
and data access for the MAC operands), specialized hardware accelerators are
very suitable for their execution. Various dedicated accelerators have been
proposed in the literature for the efficient processing of
CNNs~\cite{Jouppi2017,ChenEy2017,Bitfusion2018,shidiannao2015,NullHop2018,JSim2016}. The
architecture in Google TPU~\cite{Jouppi2017} consists of a 2D array of parallel
MAC computation units, a large on-chip buffer for the storage of ifmap and psum
data, additional local storage inside the MAC computation core for the filter
data, and finally, off-chip memory to store all the feature maps and filters
together. Similarly, along with an array of parallel processing elements, the
architectures in~\cite{JSim2016, Bitfusion2018,shidiannao2015, NullHop2018} use a separate
on-chip SRAM to store a chuck of filter, ifmap, and psum data, and an external
DRAM to completely store all the ifmap/ofmap and filter data. In
\cite{shidiannao2015,ChenEy2017}, local storage is used inside each processing
element while allowing data communication between processing elements to
facilitate better reuse of data.
Additionally,~\cite{JSim2016,NullHop2018,ChenEy2017} exploit the inherent data
sparsity in the internal layers of CNN to save computation energy.

It is evident that the key architectural features of these accelerators are
fundamentally similar: an array of processing elements to perform neural
computation in parallel, multiple levels of memory for fast data access, and
greater data reuse. We utilize these key architectural features of ASIC hardware to develop the general framework of CNNergy.

\subsubsection{General Framework of CNNergy}
\label{sec:AM_framework}
We develop CNNergy to estimate the energy dissipation in a CNN hardware accelerator. 
The general framework of CNNergy is illustrated in Fig.~\ref{fig:GenFrm}. 
We use CNNergy to determine the {\em in situ} computation energy ($E_L$
in~\eqref{eq:Ecost}), accounting for scheduling and computation overheads.

One of the largest contributors to energy is the cost of data access from
memory. Thus, data reuse is critical for energy-efficient execution of CNN
computations to reduce unnecessary high-energy 
memory accesses, particularly the ifmap and filter weights, and is used
in~\cite{Jouppi2017,ChenEy2017,shidiannao2015,NullHop2018,JSim2016}. This may
involve, for example, ifmap and filter weight reuse across convolution windows;
ifmap reuse across filters, and reduction of psum terms across channels.  Given
the accelerator memory hierarchy, number of parallel processing elements, and CNN
layer shape, Block~\circled{1} of Fig.~\ref{fig:GenFrm} is an automated scheme
for scheduling MAC computations while maximizing data reuse.  The detailed
methodology for obtaining these scheduling parameters is presented in
Section~\ref{sec:CompSc}.

Depending on the scheduling parameters, the subvolume of the ifmap to be
processed at a time is determined. Block~\circled{2} then computes the
corresponding energy for the MAC operations and associated data accesses.  The
computation in Block~\circled{2} is repeated to process the entire data volume
in a layer, as detailed in Section~\ref{sec:CompEnergy}.

The framework of CNNergy is general and its principles apply to a large class of
CNN accelerators. However, to validate the framework, we demonstrate it on a
specific platform, Eyeriss~\cite{ChenEy2017}, for which ample performance data
is available, including silicon measurements. 
Eyeriss has an array of $J \times K$ processing elements (PEs), each with:
\begin{itemize}
\item a multiply-accumulate (MAC) computation unit.
\item register files (RFs) for filter, ifmap, and psum data. 
\end{itemize}
We define $f_s$, $I_s$, and $P_s$ as the maximum number of $b_w$-bit filter,
ifmap, and psum elements that can be stored in a PE.

The accelerator consists of four levels of memory: DRAM, global SRAM buffer
(GLB), inter-PE RF access, and local RF within a PE.  During the computations
of a layer, filters are loaded from DRAM to the RF.  In the GLB, storage is
allocated for psum and ifmap. After loading data from DRAM, ifmaps are stored
into the GLB to be reused from the RF level. The irreducible psums navigate
through GLB and RF as needed. After complete processing of a 3D ifmap, the
ofmaps are written back to DRAM.

\section{Analytical CNN Energy Model}
\label{sec:AMsection}

\noindent
We formulate an analytical model (CNNergy) for the CNN processing energy (used in
\eqref{eq:Ecost}), $E_L$, up to the $L^{th}$ layer, as
\begin{eqnarray}
\textstyle E_{L} = \sum_{i=1}^{L} E_{Layer} (i)
\label{eq:Eboard}
\end{eqnarray}
where $E_{Layer}(i)$ is the energy required to process layer $i$ of the CNN.
To keep the notation compact, we drop the index ``$(i)$'' in the remainder of this
section.  We can write $E_{Layer}$ as:
\begin{eqnarray}
E_{Layer} &=& E_{Comp} + E_{Cntrl} + E_{Data} \label{eq:Elayer}
\end{eqnarray}
where $E_{Comp} $ is the energy to compute MAC operations associated with the
$i^{th}$ layer, $E_{Cntrl}$ represents the energy associated with the control and
clocking circuitry in the accelerator, and $E_{Data}$ is the memory data access energy,
\begin{eqnarray}
E_{Data} &=& E_{onChip-data} + E_{DRAM} \label{eq:Edata} 
\end{eqnarray}
i.e., the sum of data access energy from on-chip memory (from GLB,
Inter-PE, and RF), and from the off-chip DRAM.

The computation of these energy components, particularly the data access
energy, is complicated by their dependence on the data reuse pattern in the
CNN. In Sections~\ref{sec:ConceptAM} to \ref{sec:CompEnergy}, we develop a heuristic for optimal data
reuse and describe the methodology in our CNNergy for estimating these energy components.

\begin{figure}[!ht]
\vspace{-0.2cm}
\centering
\includegraphics[height=4.0cm]{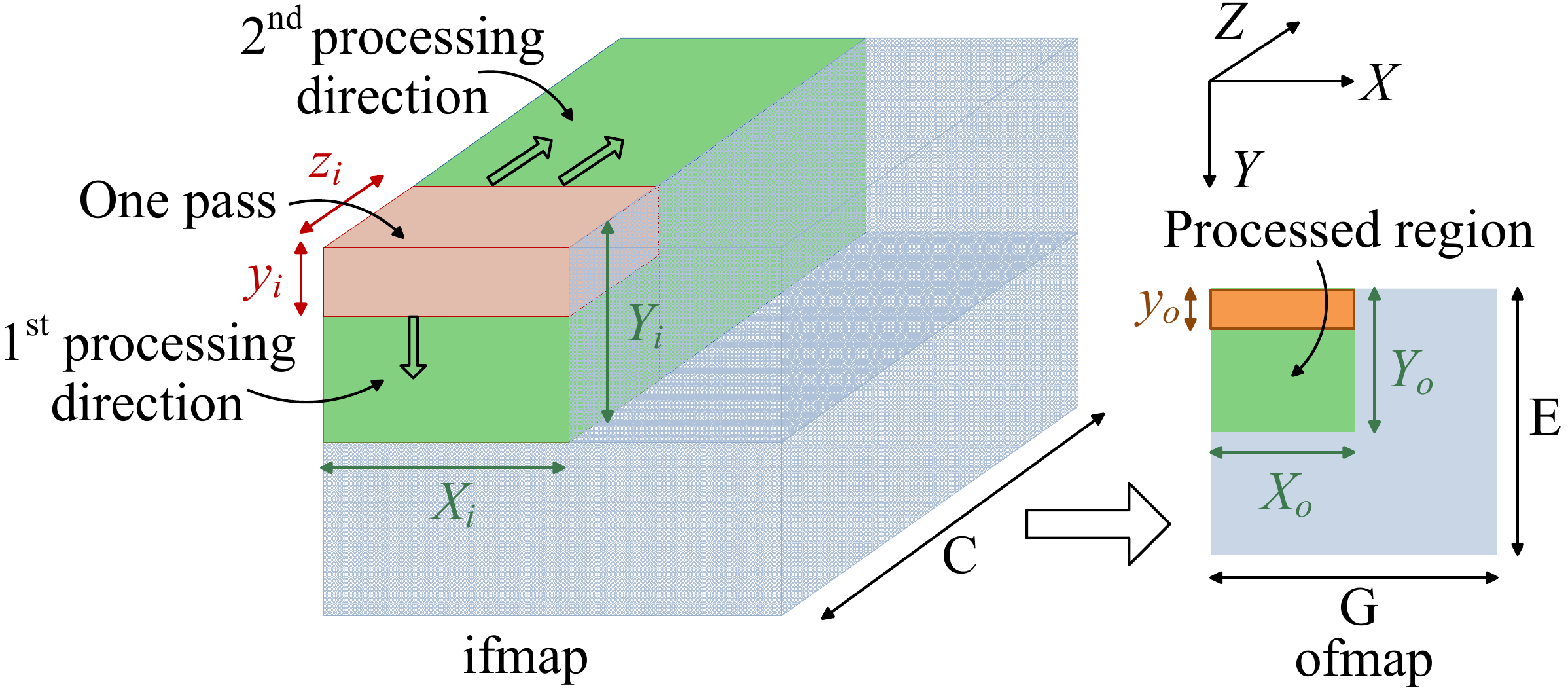}
\caption{A simplified schematic illustrating how CNNergy processes
an ifmap (for one 3D filter).}
\label{fig:3Dplot}
\vspace{-0.2cm}
\end{figure}

Specifically, in Section~\ref{sec:ConceptAM}, we conceptually demonstrate how CNNergy processes the 3D data volume by dividing it into multiple subvolumes. We also define the parameters to schedule the CNN computation in this section. Next, in Section~\ref{sec:DataFlowPE}, we describe the dataflow to distribute the convolution operation in the PE array and identify the degrees of freedom to map the computation in the hardware. In Section~\ref{sec:CompSc}, we present our automated mapping scheme to compute the computation scheduling parameters for any given CNN layer. Finally, using the scheduling parameters, we present the steps to compute energy for each component of \eqref{eq:Elayer} in Section~\ref{sec:CompEnergy}.

\subsection{Conceptual Illustration of CNNergy}
\label{sec:ConceptAM}

Fig.~\ref{fig:3Dplot} illustrates how the 3D ifmap is processed by convolving
one 3D filter with ifmap to obtain one 2D channel of ofmap; this is repeated
over all filters to obtain all channels of ofmap.  Due to the large volume of
data in a CNN layer and the limited availability of on-chip storage (register
files and SRAM), the data is divided into smaller subvolumes, each of which is
processed by the PE array in one \underline{\bf pass} to generate psums, where the
capacity of the PE array and local storage determine the amount of data that
can be processed in a pass.

All psums are accumulated to produce the final ofmap entry; if only a subset of
psums are accumultated, then the generated psums are said to be {\em
irreducible}.  The pink region of size $X_i \times y_i \times z_i$ shows the
ifmap volume that is covered in one pass, while the green region shows the
volume that is covered in multiple passes before a write-back to DRAM. 

As shown in the figure (for reasons provided in Section~\ref{sec:CompSc}),
consecutive passes first process the ifmap in the $X$-direction, and then the
$Y$-direction, and finally, the $Z$-direction.  After a pass, irreducible psums
are written back to GLB, to be later consolidated with the remainder of the
computation to build ofmap entries.  After processing the full $Z$-direction
(i.e., all the channels of a filter and ifmap) the green ofmap region of size
$X_o \times Y_o$ is formed and then written back to DRAM. The same process is
then repeated until the full volume of ifmap/ofmap is covered.

Fig.~\ref{fig:3Dplot} is a simplified illustration that shows the processing of
one 3D ifmap using one 3D filter. Depending on the amount of available register
file storage in the PE array, a convolution operation using $f_i \geq 1$ filters
can be performed in a pass. Furthermore, subvolumes from multiple
images (i.e., $N$ ifmaps) can be processed together, depending on the SRAM
storage capacity.

Due to the high cost of data fetches, it is important to optimize the pattern
of fetch operations from the DRAM, GLB, and register file by reusing the
fetched data.  The level of reuse is determined by the parameters $f_i$, $z_i$,
$y_i$, $y_o$, $X_i$, $X_o$, $Y_i$, $Y_o$, and $N$. Hence, the efficiency of the
computation is based on the choice of these parameters.  The mapping approach
that determines these parameters, in a way that attempts to minimize data
movement, is described in Section~\ref{sec:CompSc}.

\begin{figure}[!ht]
\centering
\includegraphics[width=3.5in]{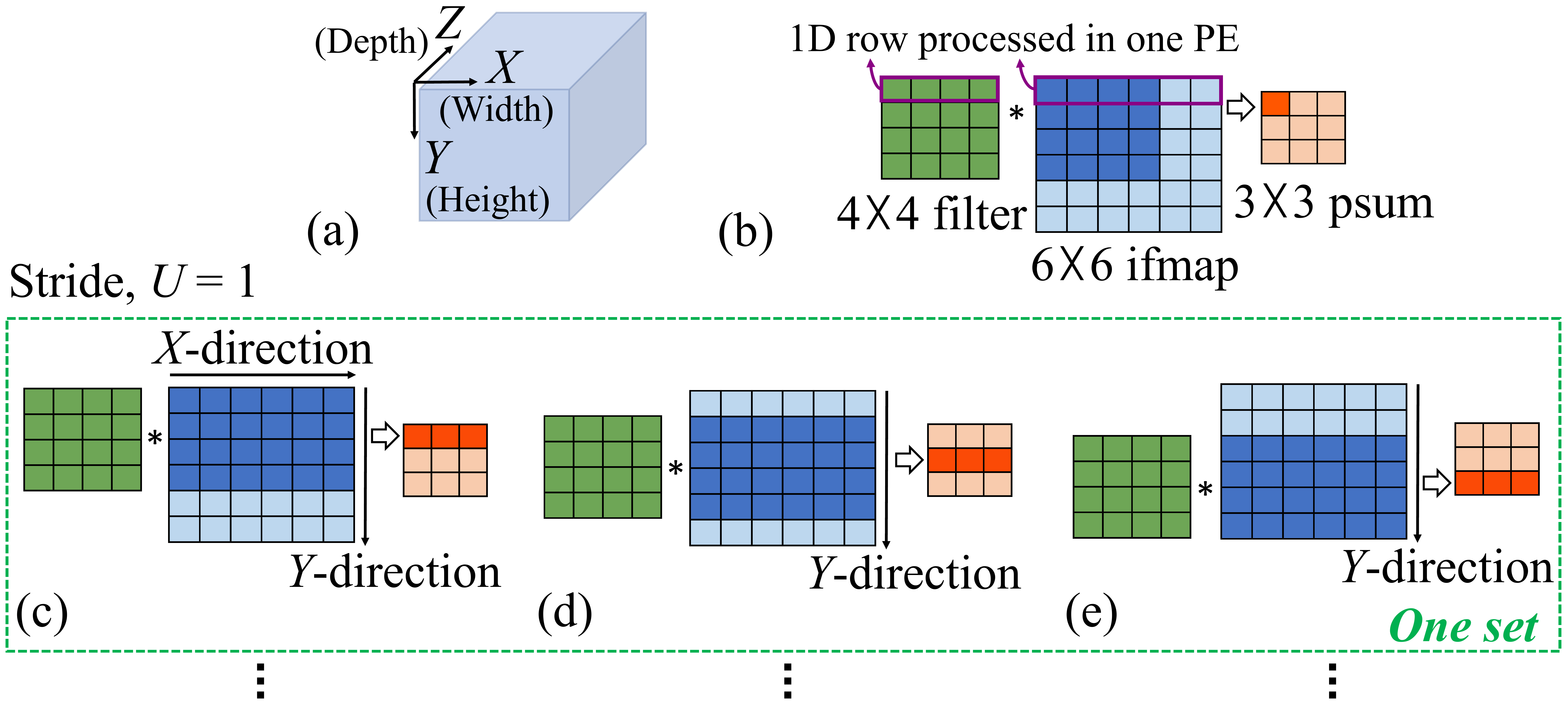}
\caption{
(a) The $X$-, $Y$-, $Z$-directions.
(b)--(e) Example showing the convolution operation in the PEs (adapted from~\cite{Sze2017}).}
\label{fig:PEEx}
\end{figure}

\subsection{Dataflow Illustration in the PE array}
\label{sec:DataFlowPE}

\noindent
In this section we describe how the convolution operations are distributed in
the 2D PE array of size $J \times K$. We use the row-stationary scheme to
manage the dataflow for convolution operations in the PE array as it is shown to offer
higher energy-efficiency than other alternatives~\cite{Sze2017,Chen2017}.

\subsubsection{Processing the ifmap in Sets -- Concept}
\label{sec:ifmap-concept}

\noindent
We explain the row-stationary dataflow with a simplified example shown in
Fig.~\ref{fig:PEEx}, where a single channel of the filter and ifmap are
processed (i.e., $z_i = 1$). Fig.~\ref{fig:PEEx}(b) shows a basic 
computation where a $4 \times 4$ filter (green region) is multiplied with the
part of the ifmap that it is overlaid on, shown by the dark blue region.  Based
on the row-stationary scheme for the distributed computation, these four rows
of the ifmap are processed in four successive PEs within a column of the
PE array.  Each PE performs an element-wise multiplication of the ifmap row and
the filter row to create a psum.  The four psums generated are transmitted to
the uppermost PEs and accumulated to generate their psum (dark orange).

Extending this operation to a full convolution implies that the ifmap slides
under the filter in the negative $X$-direction with a stride of $U$, while
keeping the filter stationary: for $U=1$, two strides are required to cover the
ifmap in the $X$-direction. In our example, for each stride, each of the four
PEs performs the element-wise multiplication between one filter row and one
ifmap row, producing one 1D row of psum, which is then accumulated to produce
the first row of psum, as illustrated in Fig.~\ref{fig:PEEx}(c). 

Thus, the four dark blue rows in Fig.~\ref{fig:PEEx}(c) are processed by four
PEs (one per row) in a column of the PE array.  The reuse of the filter avoids
memory overheads due to repeated fetches from other levels of memory.
To compute psums associated with other rows of the ofmap, a subarray of 12 PEs
(4 rows $\times$ 3 columns) processes the ifmap under a $Y$-direction ifmap
stride.  The ifmap regions thus processed are shown by the dark regions of
Fig.~\ref{fig:PEEx}(d),(e).

We define the amount of processing performed in $R$ rows, across all $K$
columns of the PE array, as a \underline{\bf \em set}. For a $4 \times 3$ PE array, a set
coresponds to the processing in Fig.~\ref{fig:PEEx}(c)--(e).

In the general context of the $J \times K$ PE array, a set is formed from $R
\times K$ PEs.  Therefore, the number of sets which can fit in the full PE
array (i.e., the number of sets in a pass) is given by:
\begin{eqnarray}
S_{Pass} = \left \lfloor \frac{J \times K}{R \times K} \right \rfloor = 
\left \lfloor \frac{J}{R} \right \rfloor,
\label{eq:SetEq}
\end{eqnarray}
i.e., $S_{Pass}$ is the ratio of the PE array height to the filter height.

\subsubsection{Processing the ifmap in Sets -- Realistic Scenario}
\label{sec:ifmap-realistic}

\noindent
We now generalize the previously simplified assumptions in the example to
consider typical parameter ranges.  We also move from the assumption of
$z_i = 1$, to a more typical $z_i > 1$.

First, the filter height can often be less than the size of the PE
array.  When $R < J$, the remaining PE array rows can process more
filter channels simultaneously, in multiple sets.

Second, typical RF sizes in each PE are large enough to operate on more than
one 1D row.  Under this scenario, {\em within each set}, a group of 2D filter planes
is processed.  There are several degrees of freedom in mapping computations to
the PE array.  When several 1D rows are processed in a PE, the
alternatives for choosing this group of 1D rows include:
\begin{enumerate}
\item[(i)] choosing filter/ifmap rows from different channels
of the same 3D filter/ifmap;
\item[(ii)] choosing filter rows from different 3D filters;
\item[(iii)] combining (i) and (ii), where some rows come from the channels of
the same 3D filter and some rows come from the channels under different 3D
filters.
\end{enumerate}

{\em Across sets} in the PE array, similar mapping choices are available.
Different groups of filter planes (i.e., channels) are processed in different
sets. These groups of planes can be chosen either from the same 3D filter, or
from different 3D filters, or from a combination of both. 

Thus, there is a wide space of mapping choices for performing the CNN
computation in the PE array.

\subsubsection{Data Reuse}
\label{sec:data-reuse}

\noindent
Due to the high cost of data accesses from the next memory level,  it is
critical to reuse data as often as possible to achieve energy efficiency.
Specifically, after fetching data from a higher access-cost memory level, data
reuse refers to the use of that data element in multiple MAC operations. For
example, after fetching a data element from GLB to RF, if that data is used
across $r$ MAC operations, then the data is reused $r$ times with respect to
the GLB level.

We now examine the data reuse pattern within the PE array.  Within each PE
column, an ifmap plane is being processed along the $X$-direction, and multiple
PE columns process the ifmap plane along the $Y$-direction.  Two instances of data
reuse in the PE array are:
\begin{enumerate}
\item[(1)] In each set, the {\em same ifmap row} is processed along the PEs in a
diagonal of the set.  This can be seen in the example set in
Fig.~\ref{fig:PEEx}(c)-(e), where the third row of the ifmap plane is common in
the PEs in $r_3 c_1$ in (c), $r_2 c_2$ in (d), and $r_1 c_3$ in (e), where $r_i
c_j$ refers to the PE in row $i$ and column $j$.
\item[(2)] The {\em same filter row} is processed in the PEs in a row: in
Fig.~\ref{fig:PEEx}(c)-(e), the first row of the filter plane is common to all
PEs in all three columns of row 1.
\end{enumerate}

Thus, data reuse can be enabled by broadcasting the same ifmap data (for
instance (1)) and the same filter data (for instance (2)) to multiple PEs for
MAC operations after they are fetched from a higher memory level (e.g., DRAM or
GLB).

\begin{table}[htb]
\centering
\caption{List of parameters for computation scheduling and accelerator hardware constraints.}
\label{tbl:NotParam}
\begin{tabular}{|c|l|}
\hline
Notation & \multicolumn{1}{c|}{Description}                                                                        \\ \hline
\multicolumn{2}{|c|}{\textbf{Computation Scheduling Parameters}}                                                   \\ \hline
$f_i$       & \#of filters processed in a pass                                                                        \\ \hline
$z_i$       & \#of ifmap/filter channels processed in a pass                                                          \\ \hline
$y_i$ ($y_o$)  & Height of ifmap (ofmap) processed in a pass                                                             \\ \hline
$X_i$ ($X_o$)  & Width of ifmap (ofmap) processed in a pass                                                              \\ \hline
$Y_i$ ($Y_o$)  & \begin{tabular}[c]{@{}l@{}}Height of ifmap (ofmap) processed before a\\ write back to DRAM\end{tabular} \\ \hline
$N$        & \#of ifmap from different images processed together                                                     \\ \hline
\multicolumn{2}{|c|}{\textbf{Accelerator Hardware Parameters}}                                                     \\ \hline
$f_s$       & Size of RF storage for filter in one PE                                                                 \\ \hline
$I_s$       & Size of RF storage for ifmap in one PE                                                                  \\ \hline
$P_s$       & Size of RF storage for psum in one PE                                                                   \\ \hline
$J$         & Height of the PE array (\#of rows)                                                                      \\ \hline
$K$         & Width of the PE array (\#of columns)                                                                    \\ \hline
$|$GLB$|$   & Size of GLB storage                                                                                     \\ \hline
$b_w$       & bit width of each data element	
              \\ \hline
\end{tabular}
\vspace{-0.4cm}
\end{table}

\subsection{Obtaining Computation Scheduling Parameters}
\label{sec:CompSc}

\noindent
As seen in Section~\ref{sec:DataFlowPE}, depending on the specific CNN and its
layer structure, there is a wide space of choices for computations to be
mapped to the PE array.  The mapping of filter and ifmap parameters to the PE
array varies with the CNN and with each layer of a CNN. This mapping is a
critical issue in ensuring low energy, and therefore, in this work, we develop
an automated mapping scheme for any CNN topology.  The scheme computes the
parameters for scheduling computations. The parameters are described in
Section~\ref{sec:ConceptAM} and summarized in Table~\ref{tbl:NotParam}. The
table also includes the parameters for the accelerator hardware constraints.

For general CNNs, for each layer, we develop a mapping strategy that follows
predefined rules to determine the computation scheduling. The goal of
scheduling is to attempt to minimize the movement of three types of data (i.e.,
ifmap, psum, and filter), since data movement incurs large energy overheads.
In each pass, the mapping strategy uses the following priority rules:\\
{\em (i)} We process the maximum possible channels of an ifmap to reduce
the number of psum terms that must move back and forth with the next level of memory.\\
{\em (ii)} We prioritize filter reuse, psum reduction over ifmap reuse.

The rationale for Rules {\em (i)} and {\em (ii)} is that since a very large
number of psums are generated in each layer, psum reduction is the most
important factor for energy, particularly because transferring psums to the
next pass involves expensive transactions with the next level of memory.
This in turn implies that filter weights must
remain stationary for maximal filter reuse.  Criterion {\em (ii)} lowers the
number of irreducible psums: if the filter is changed and ifmap is kept fixed,
the generated psums are not reducible.

In processing the ifmap, proceeding along the $X$- and $Y$-directions enables
the possibility of filter reuse as the filter is kept stationary in the RF
while the ifmap is moved. In contrast, if passes were to proceed along the
$Z$-direction, filter reuse would not be possible since new filter channels
must be loaded from the DRAM for the convolution with ifmap.  Therefore, the
$Z$-direction is the last to be processed.  In terms of filter reuse, the $X$-
and $Y$-directions are equivalent, and we arbitrarily prioritize the
$X$-direction over the $Y$-direction.

We use the notion of a set and a pass (Section~\ref{sec:DataFlowPE}) in the
flow graph to devise the choice of scheduling parameters:

\subsubsection{Computing $y_i$ and $y_o$}
The value of $y_o = \min(K,E)$ and is limited by the number of columns, $K$, in
the PE array.  The corresponding value of $y_i$ is found using the relation 
\begin{eqnarray}
y_o = (y_i - R)/U+1
\label{Newyo}
\end{eqnarray}

\subsubsection{Computing $z_i$ and $f_i$}

The number of channels of each ifmap in a pass is computed as 
\begin{eqnarray}
z_i &=& C_{Set} \times S_{Pass}, \hspace{0.1in} 
\label{eq:AM2a} 
\end{eqnarray}
where $C_{Set}$ is the number of channels per set, and $S_{Pass}$ is the
number of sets per pass (given by~\eqref{eq:SetEq}).  Recall that the first
priority rule of CNNergy is to process the largest possible number of
ifmap channels at a time.  Therefore, to compute $C_{Set}$, we find the number
of filter rows that can fit into an ifmap RF, i.e., $C_{Set} = \lfloor I_{s}/S \rfloor$.

To enable per-channel convolution, the filter RF of a PE must be loaded with the
same number of channels as $I_s$ from a single filter.  The remainder of the dedicated
filter RF storage can be used to load channels from different filters
so that one ifmap can be convolved with multiple filters resulting in ifmap
reuse. Thus, after maximizing the number of channels of an ifmap/filter to be
processed in a pass, the remaining $f_s$ storage can be used to enable ifmap
reuse. Therefore, the number of filters processed in a pass is
\begin{eqnarray}
f_i = \lfloor f_s/I_s \rfloor
\label{eq:AM2b} 
\end{eqnarray}

\subsubsection{Computing $X_i$, $X_o$, $Y_i$, $Y_o$, $N$}

During a pass, the ifmap corresponds to the pink region in
Fig.~\ref{fig:3Dplot}, and over multiple passes, the entire green volume of the
ifmap in the figure is processed before a writeback to DRAM.  

We first compute $|$ifmap$|$ and $|$psum$|$, the storage requirements of ifmap
and psum, respectively, during the computation. The pink region has dimension $X_i
\times y_i \times z_i$ and over several passes it creates, for each of the $f_i$ filters, a set of
psums for the $X_o \times Y_o$ region of the ofmap that are not fully reduced
(i.e., they await the results of more passes). Therefore,
\begin{align}
\mbox{$|$ifmap$|$} =& \; b_w ( X_i \times y_i \times z_i ) 
\label{eq:AM3} \\
\mbox{$|$psum$|$}  =& \; b_w ( X_o \times Y_o \times f_i )
\label{eq:AM4}
\end{align}
where $b_w$ corresponds to the bit width for ifmap and psum.

Next, we determine how many ifmap passes can be processed for a limited GLB
size, $|$GLB$|$.  This is the number, $N$, of pink regions that can fit within
the GLB, i.e.,
\begin{eqnarray}
N = \left \lfloor \frac{\mbox{$|$GLB$|$}}{\mbox{$|$ifmap$|$} + \mbox{$|$psum$|$}} \right \rfloor
\label{eq:Ndefinition}
\end{eqnarray}

To compute $X_i$, we first set it to the full ifmap width, $W$, and we set
$Y_o$ to the full ofmap height, $E$, to obtain $N$.  If $N = 0$, i.e.,
$|$ifmap$|+|$psum$|>|$GLB$|$, then $X_i$ and $Y_o$ are reduced until the data
fits into the GLB and $N \geq 1$.

From the values of $X_i$ and $Y_o$ computed above, we can determine $X_o$ and
$Y_i$ using the relations 
\begin{align}
X_o =& \; \frac{X_i - S}{U} + 1 \; \; ; \; \;
Y_o = \; \frac{Y_i - R}{U} + 1
\label{eq:AM5}
\end{align}

\begin{figure}[!ht]
\centering
\includegraphics[height=9.5cm]{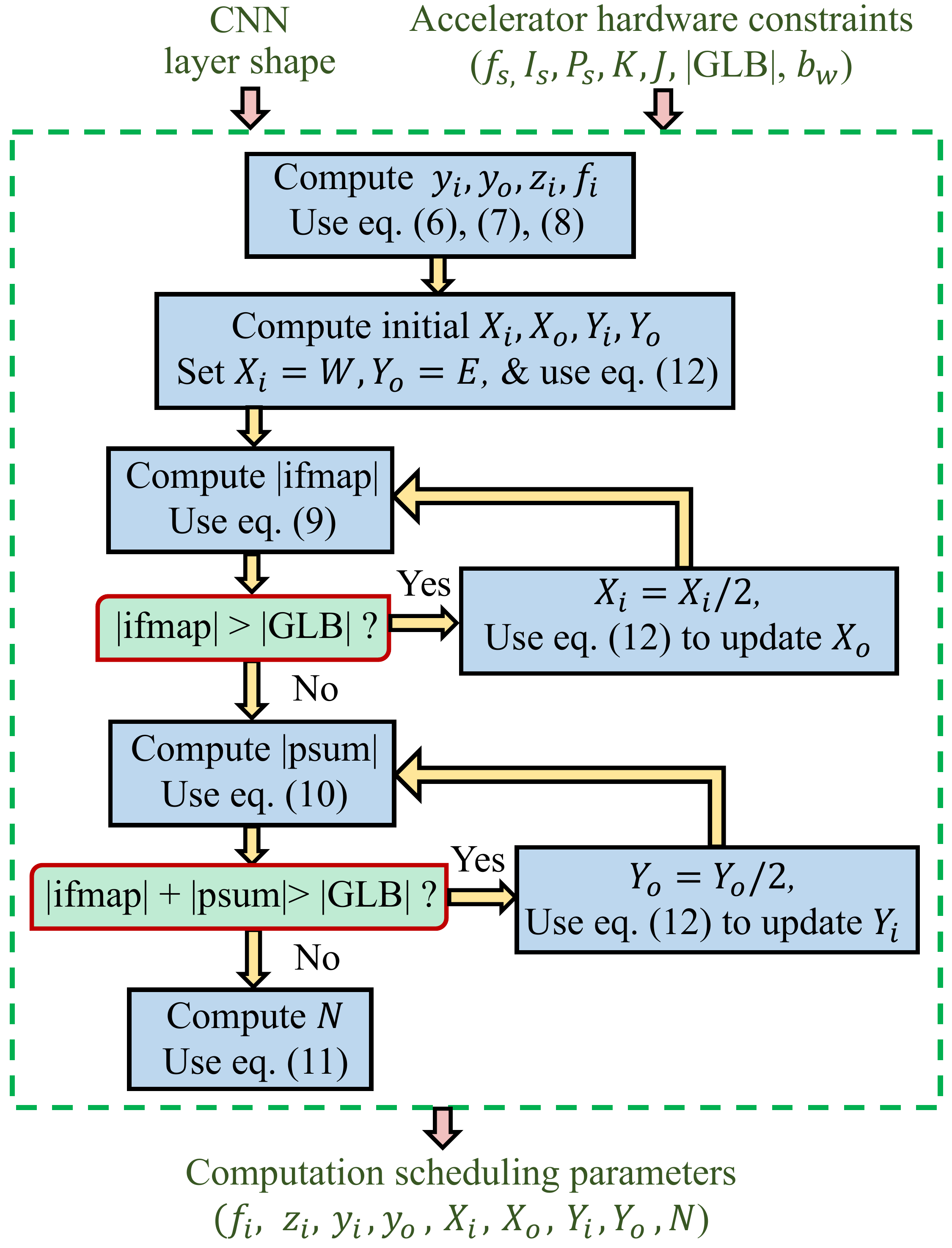}
\caption{Flow graph to obtain the computation scheduling parameters.}
\label{fig:MapFlow}
\vspace{-0.4cm}
\end{figure}

Fig.~\ref{fig:MapFlow} shows the flow graph that summarizes how the parameters
for scheduling the CNN computation are computed. The module takes the CNN layer
shape (Table~\ref{tbl:CNNParam}) and the accelerator hardware parameters
(Table~\ref{tbl:NotParam}) as inputs.  Based on our automated mapping strategy,
the module outputs the computation scheduling parameters
(Table~\ref{tbl:NotParam}). 

\subsubsection{Exception Rules} 
The mapping method handles exceptions:
\begin{itemize}
\item
If $Y_o < y_o$, some PE columns will remain unused. This is avoided by setting
$Y_o = y_o$. If the new $|$ifmap$|+|$psum$|>|$GLB$|$, $f_i$ is reduced
so that the data fits into the GLB.
\item
If $C < z_i$, all channels are processed in a pass while increasing $f_i$,
as there is more room in the PE array to process more filters. The
cases $F < f_i$, $P_s < f_i$ proceed by reducing $f_i$.
\item
All Conv layers whose filter has the dimension $R = S = 1$ 
(e.g., inside the inception modules of GoogleNet, or the fire modules of
SqueezeNet) are handled under a fixed exception rule that uses a reduced $z_i$,
and suitably increased $f_i$.
\end{itemize}
The exceptions are triggered only for a few CNN layers (i.e., layers having
relatively few channels or filters).

\begin{algorithm}[h]
{\scriptsize
\textbf{INPUT:}  
Computation scheduling parameters: $f_i$, $z_i$, $y_i$, $y_o$, $X_i$, $X_o$, $Y_i$, $Y_o$, $N$
(defined in Table~\ref{tbl:NotParam});\\
CNN layer shape parameters: $R$, $S$, $H$, $W$, $E$, $G$, $C$, $F$ (defined in Table~\ref{tbl:CNNParam});\\
Technology parameters for energy per operation (specified in Table~\ref{tbl:TechP})\\
\textbf{OUTPUT:} 
Energy to process a CNN layer, $E_{Layer}$
\\
\textbf{STEPS:}
\begin{algorithmic} [1]
\State 
Determine the subvolume from each ifmap ($X_i \times y_i \times z_i$) and from
each filter ($R \times S \times z_i$) processed in one pass.
\label{alg:Ln1}
\State 
Determine the number of psums ($X_o \times y_o$) generated during a pass from
each 3D filter and each image.
\label{alg:Ln2}
\State 
Compute $I_{Pass}$, \# of ifmap elements accessed in one pass from DRAM and
GLB, using~\eqref{eq:AM7} 
\label{alg:Ln3}
\State
Compute $P_{Pass}$, \# of psum elements to read (write) from (to) GLB during one
pass, using~\eqref{eq:AM8}
\label{alg:Ln4}
\State 
Compute $F_{Pass}$, \# of filter elements to load from DRAM for one pass,
using~\eqref{eq:AM6} 
\label{alg:Ln5}
\State
Compute \# of passes before a writeback of ofmap to DRAM ($\frac{Y_o}{y_o}
\times \frac{C}{z_i}$ passes along $Y$ and $Z$-directions)
\label{alg:Ln6}
\State
Compute $E_{D_{X_iY_iz_i}}$, data access energy to process a subvolume of ifmap
over which filter data is reused, using~\eqref{eq:AM9}
\label{alg:Ln7}
\State
Repeat Step~\ref{alg:Ln7} to compute $E_{D_{X_o Y_o}}$, data access energy to
produce and perform DRAM write of $X_o \times Y_o$ region of each ofmap channel
over $f_i$ filters and $N$ images, using~\eqref{eq:AM9new}
\label{alg:Ln8}
\State
Repeat Step~\ref{alg:Ln8} to compute $E_{Data}$, total data access energy, using~\eqref{eq:AM10}
\label{alg:Ln9}
\State
Compute $E_{Comp}$ using~\eqref{eq:AMnew}
\label{alg:Ln10}
\State
Compute $E_{Cntrl}$ using~\eqref{eq:CN1}
\label{alg:Ln11}
\State
Compute $E_{Layer}$, using~\eqref{eq:Elayer}
\label{alg:Ln12}
\end{algorithmic}
\caption{Algorithm for Energy Computation.}
\label{alg:EnCmpAlgo}
}
\end{algorithm}

\subsection{Energy ($E_{Layer}$) Computation}
\label{sec:CompEnergy}

\noindent
In Section~\ref{sec:CompSc}, we have determined the subvolume of ifmap and filter
data to be processed in a pass.  From the scheduling parameters we can also compute the
number of passes before a writeback of ofmap to DRAM. Therefore, we have
determined the schedule of computations to generate all channels of ofmap. We
now estimate each component of $E_{Layer}$ in~\eqref{eq:Elayer}. The steps for
this energy computation are summarized in Algorithm~\ref{alg:EnCmpAlgo}, which
takes as input the computation scheduling parameters, CNN layer shape parameters,
and technology-dependent parameters that specify the energy per operation
(Table~\ref{tbl:TechP}), and outputs $E_{Layer}$.

\subsubsection{Computing $E_{Data}$, $E_{Comp}$} 

We begin by computing the subvolume of data loaded in each pass
(Lines~\ref{alg:Ln1}--\ref{alg:Ln5}).  In Fig~\ref{fig:3Dplot}, $I_{Pass}$ is
illustrated as the pink ifmap region which is processed in one pass for an
image, and $P_{Pass}$ is the number of psum entries associated with the orange
ofmap region for a single filter and single image.  The filter data is reused
across ($Y_o/y_o$) passes, and we denote the number of filter elements loaded
for these passes by $F_{Pass}$. Thus, for $f_i$ filters and $N$ images,
\begin{eqnarray}
I_{Pass} &=& N \times (X_i \times y_i \times z_i)
\label{eq:AM7} \\
P_{Pass} &=& N \times (X_o \times y_o) \times f_i
\label{eq:AM8} \\
F_{Pass} &=& f_i \times (R \times S \times z_i)
\label{eq:AM6}
\end{eqnarray}

To compute energy, we first determine $E_{D_{X_iY_iz_i}}$, the data access
energy required to process $X_i \times Y_i \times z_i$ volume of each ifmap
over $f_i$ filters and $N$ images.  In each pass, a volume $I_{Pass}$ of the
ifmap is brought from the DRAM to the GLB for data access; $P_{Pass}$ psums
move between GLB and RF; and RF-level data accesses ($RF_{MAC}$) occur for the
four operands associated with each MAC operation in a pass.  Therefore, the
corresponding energy can be computed as:

{\footnotesize
\begin{eqnarray}
E_{D_{X_iY_iz_i}} = \bigg[e_{DRAM}(I_{Pass}) + e_{GLB}(I_{Pass}) + e_{GLB}(P_{Pass}) \nonumber \\
+ e_{RF}(RF_{MAC})\bigg] \times \frac{Y_o}{y_o} + e_{DRAM}(F_{Pass})
\label{eq:AM9}
\end{eqnarray}
}

\noindent
Here, $e_{\cal O}()$ denotes the energy associated with operation ${\cal O}$,
and each energy component can be computed by multiplying the energy per
operation by the number of operations. Since filter data is reused across
($Y_o/y_o$) passes, all components in \eqref{eq:AM9}, except the energy
associated with filter access, are multiplied by this factor. Each psum is
written once and read once, and $e_{GLB}(P_{Pass})$ accounts for both
operations.

\begin{figure}[htb]
\centering
\subfigure[]{
\includegraphics[height=3.5cm]{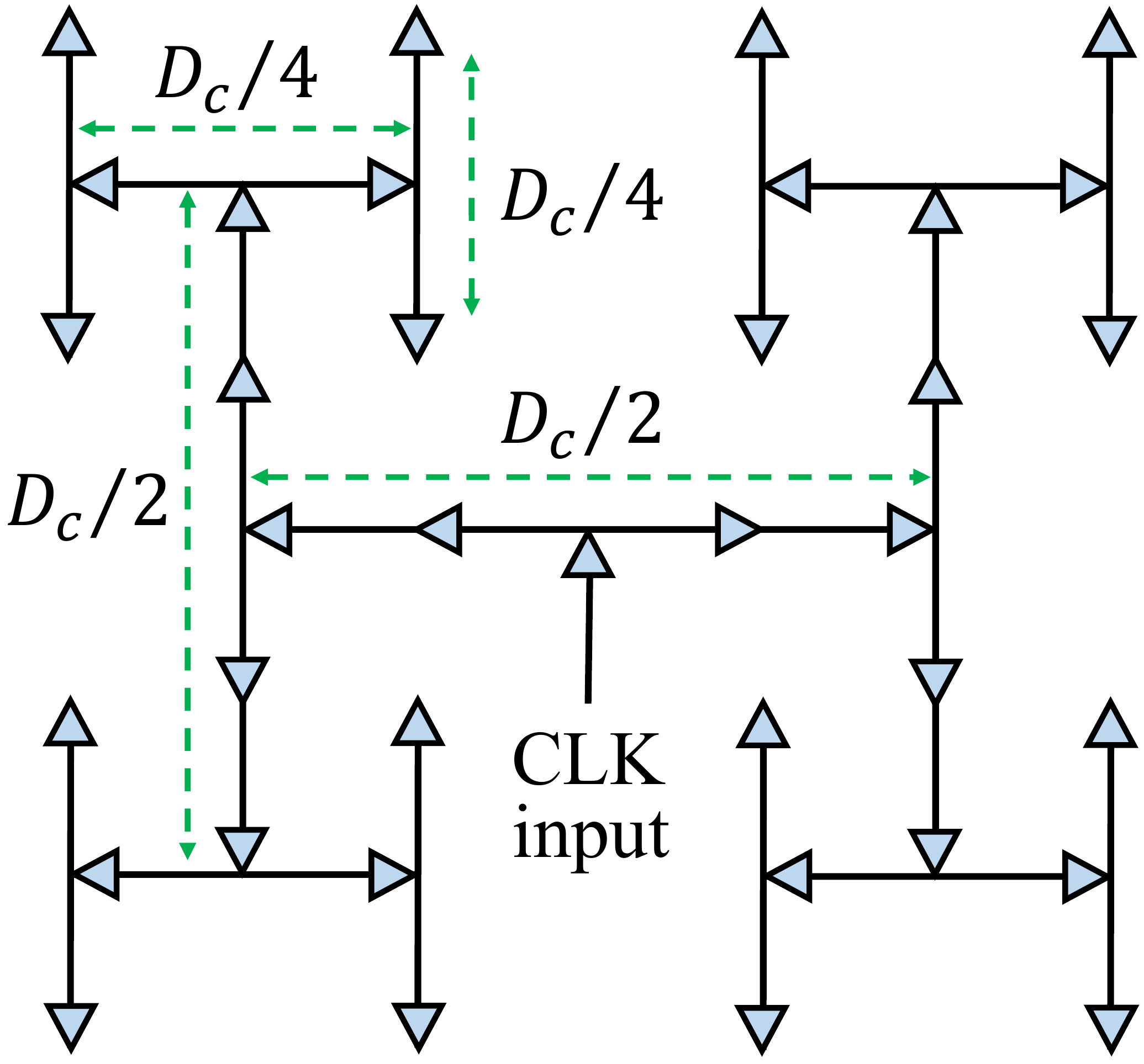}
\label{fig:HTree}
}
\subfigure[]{
\includegraphics[height=3.5cm]{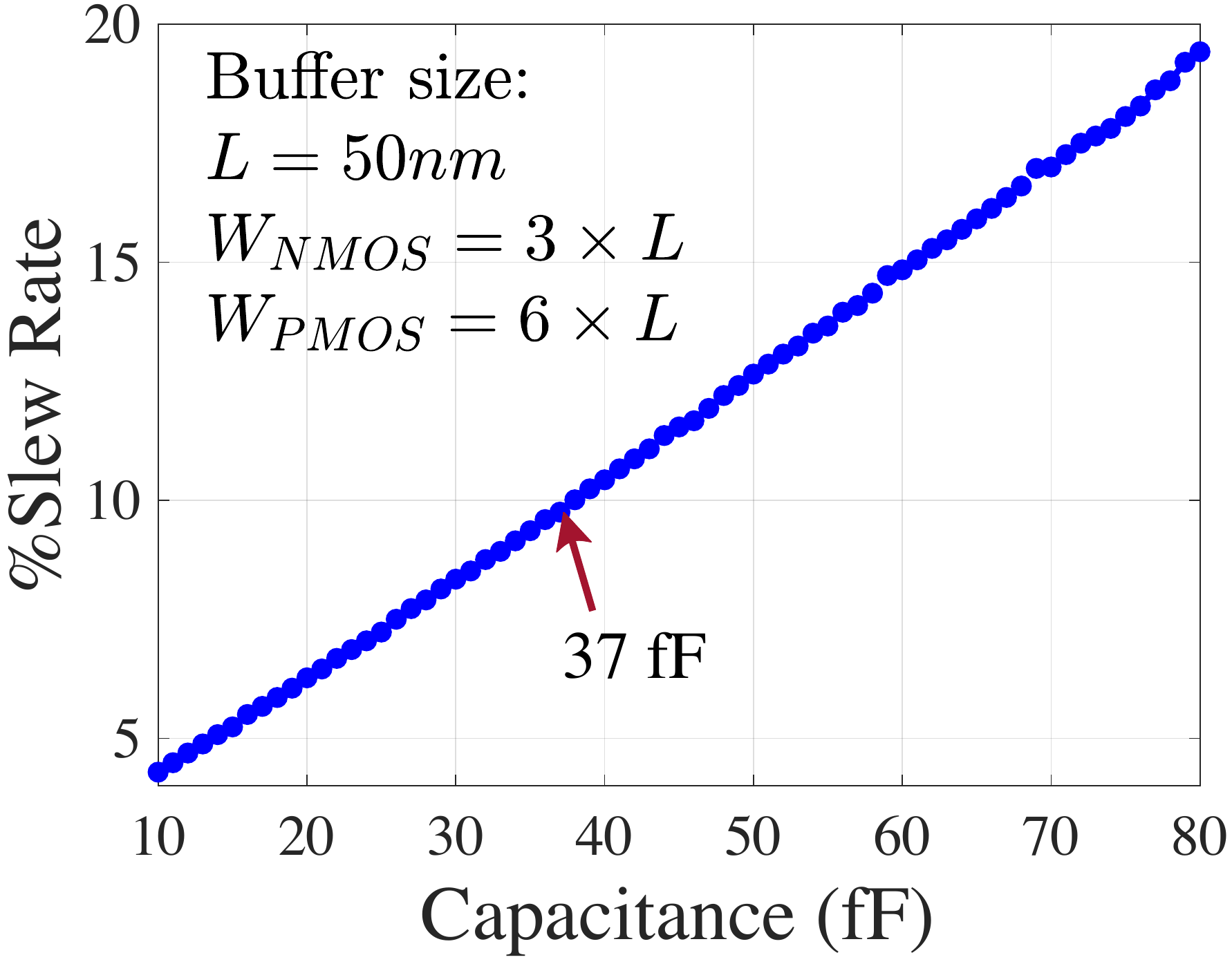}
\label{fig:SlewRate}
}
\caption{(a) H-tree style clock distribution network. (b) Percent slew of the clock vs. load capacitance driven by each stage of a clock buffer.}
\vspace{-0.3cm}
\end{figure}

Next, all $C$ channels of ifmap (i.e., the entire green ifmap region in
Fig.~\ref{fig:3Dplot}) are processed to form the green $X_o \times Y_o$ region
of each ofmap channel, and this data is written back to DRAM. To this end, we
compute $E_{D_{X_oY_o}}$, the data access energy to produce $X_o \times Y_o$
fraction of each ofmap channel over $f_i$ filters and $N$ images, by repeating
the operations in \eqref{eq:AM9} to cover all the channels:
\begin{eqnarray}
E_{D_{X_oY_o}} = \bigg( E_{D_{X_iY_iz_i}} \times \frac{C}{z_i} \bigg) + e_{DRAM}(ofmap)
\label{eq:AM9new}
\end{eqnarray}
Finally, the computation in~\eqref{eq:AM9new} is repeated to produce the entire
$G \times E$ volume of the ofmap over all $F$ filters. Therefore, the total
energy for data access is
\begin{eqnarray}
E_{Data} = E_{D_{X_oY_o}} \times \frac{G}{X_o} \times \frac{E}{Y_o} \times \frac{F}{f_i} 
\label{eq:AM10}
\end{eqnarray}
Here, the multipliers $(G/X_o)$, $(E/Y_o)$, and $(F/f_i)$ represent the number
of iterations of this procedure to cover the entire ofmap. These steps are
summarized in Lines~\ref{alg:Ln6}--\ref{alg:Ln9} of
Algorithm~\ref{alg:EnCmpAlgo}. Finally, the computation energy of the
Conv layer is computed by:
\begin{eqnarray}
E_{Comp} = N \times (R \cdot S \cdot C) \times (E \cdot G \cdot F) \times \tilde{e}_{MAC}
\label{eq:AMnew}
\end{eqnarray}
where $\tilde{e}_{MAC}$ is the energy per MAC operation, and it is multiplied by
the number of MACs required for a CNN layer.

\subsubsection{Sparsity} 

The analytical model exploits sparsity in the data (i.e., zeros in ifmap/ofmap)
at internal layers of a CNN.  Except the input ifmap to the first Conv layer of
a CNN, all data communication with the DRAM (i.e., ifmap read or ofmap write)
is performed in run-length compressed (RLC) format~\cite{ChenEy2017}. In
addition, for a zero-valued ifmap, the MAC computation as well as the
associated filter and psum read (write) from (to) RF level is skipped to reduce
energy.

\begin{figure*}[htb]
\centering
\subfigure[]{
\includegraphics[height=4.5cm]{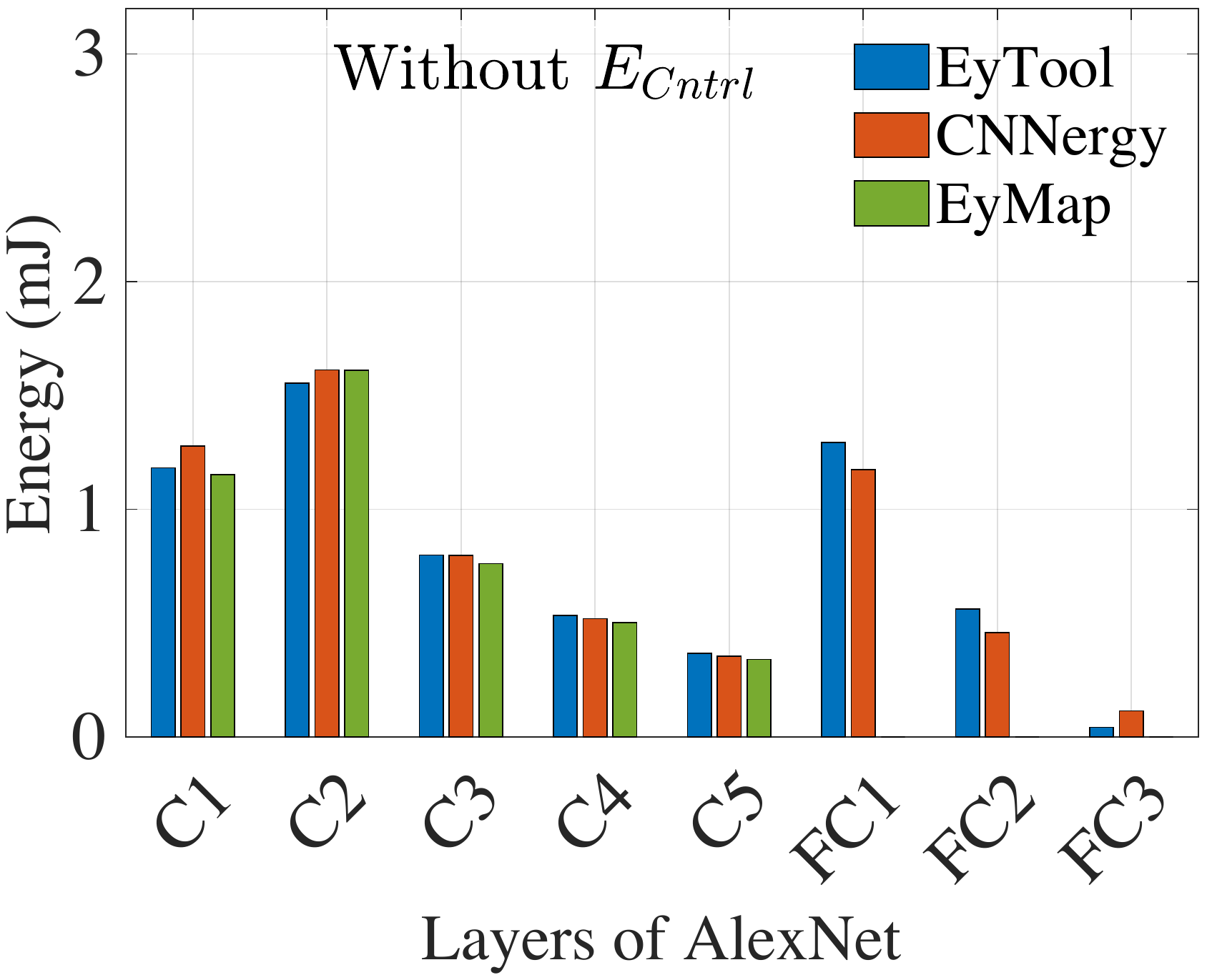}
\label{fig:AMval}
}
\subfigure[]{
\includegraphics[height=4.5cm]{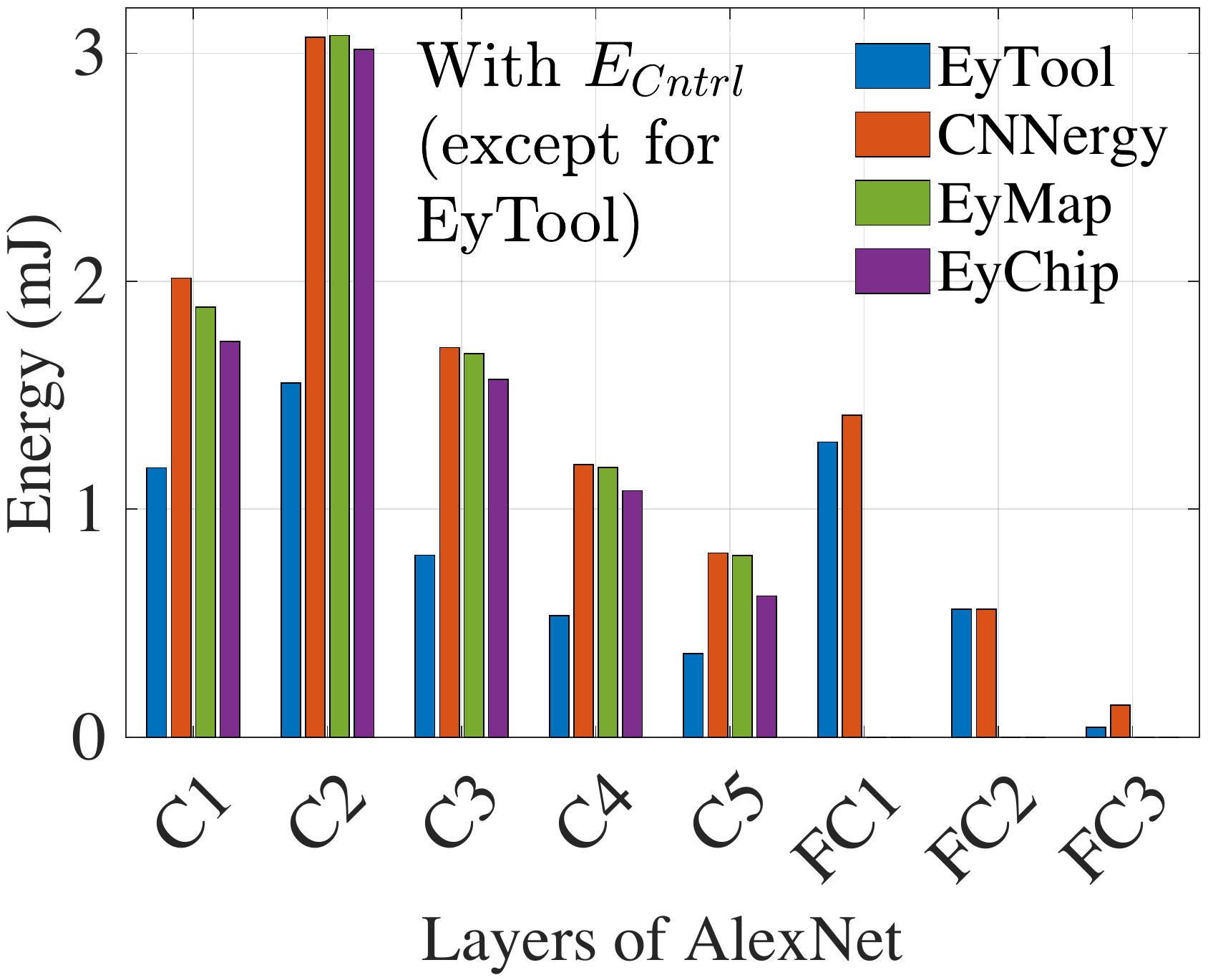}
\label{fig:Wcntrl}
}
\subfigure[]{
\includegraphics[height=4.5cm]{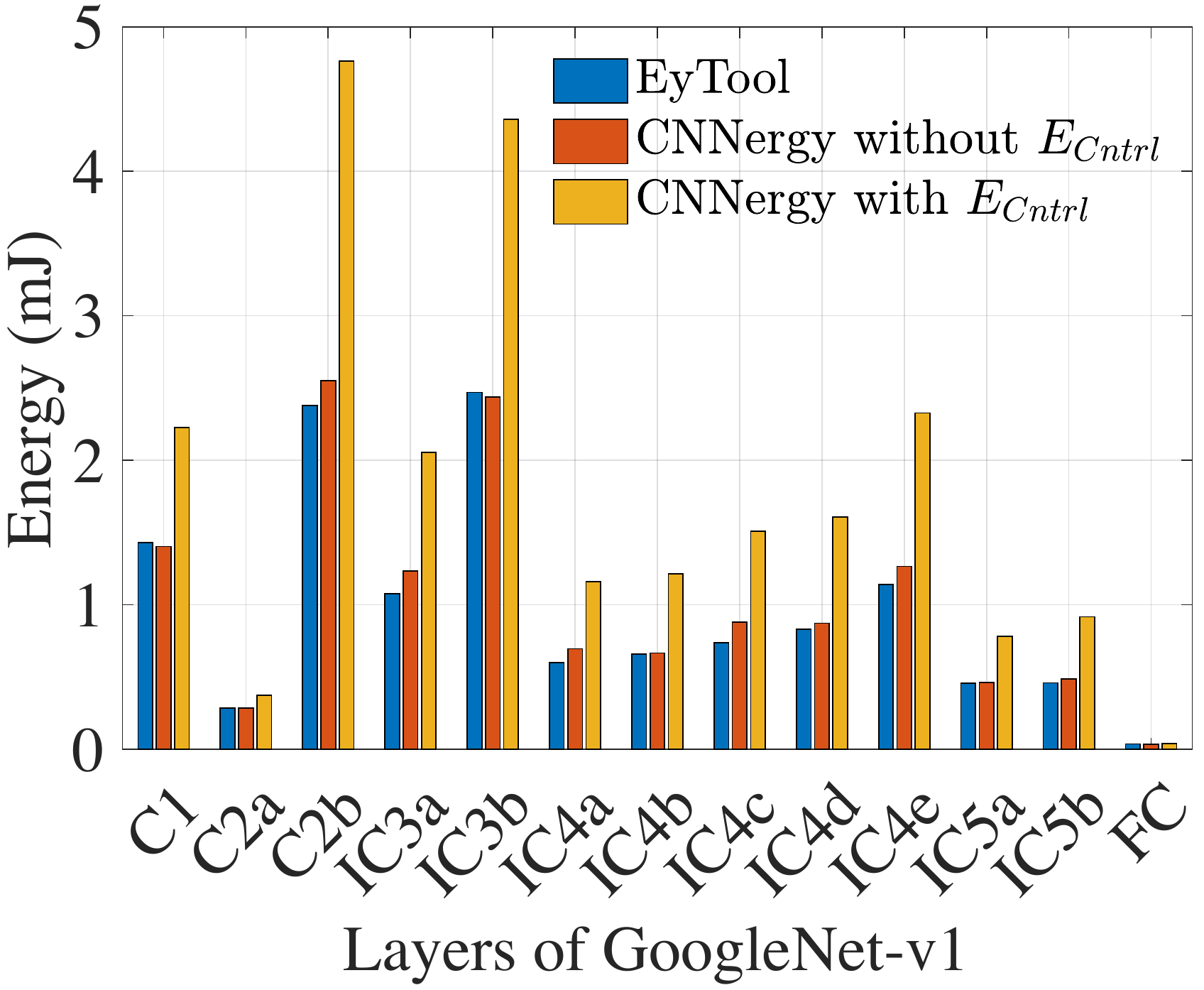}
\label{fig:AMGnet}
}
\caption{Energy validation of CNNergy: (a) AlexNet without $E_{Cntrl}$ (b) AlexNet with $E_{Cntrl}$ model (c) GoogleNet-v1.}
\label{fig:AMAll}
\vspace{-0.2cm}
\end{figure*}

\subsubsection{Computing $E_{Cntrl}$} 

The control overheard includes the clock power, overheads for control circuitry
for the PE array, network-on-chip to manage data delivery, I/O pads, etc.  Of
these, the clock power is a major contributor (documented as $\sim$33\%--45\%
in~\cite{ChenEy2017}), and other components are relatively
modest.  The total control energy, $E_{Cntrl}$, is modeled as:
\begin{eqnarray}
E_{Cntrl}  = P_{clk} \times \mbox{{\em latency}} \times T_{clk} + E_{other-Cntrl}
\label{eq:CN1}
\end{eqnarray}
where $P_{clk}$ is the clock power, {\em latency} is the number of cycles required
to process a single layer, and $T_{clk}$ is the clock period; $E_{other-Cntrl}$
is the control energy from components other than the clock network. 
We adopt similar strategy as in CACTI~\cite{Wilton1996} and ORION~\cite{ORION2009} to model the power of the clock network.
For a supply voltage of $V_{DD}$, the clock power is computed as:

\ignore{
{\footnotesize
\begin{align}
P_{clk} {}& = C_{clk} \times (V_{DD})^2 / T_{clk} + L_{clk}
\label{eq:CN2}\\
C_{clk} {}& = C_{total-wire} + C_{total-buff} + C_{total-PEreg} + C_{SRAM}
\label{eq:CN3}
\end{align}
}
}

{\footnotesize
\begin{eqnarray}
\hspace{-0.4cm}P_{clk} \hspace{-0.2cm}&=&\hspace{-0.2cm} C_{clk} \times (V_{DD})^2 / T_{clk} + L_{clk}
\label{eq:CN2}\\
\hspace{-0.4cm}C_{clk} \hspace{-0.2cm}&=&\hspace{-0.2cm} C_{total-wire} + C_{total-buff} + C_{total-PEreg} + C_{SRAM}
\label{eq:CN3}
\end{eqnarray}
}

\noindent
where $L_{clk}$ is leakage in the clock network.  The switching capacitance,
$C_{clk}$, includes the capacitances of the global clock buffers, wires, clocked
registers in the PEs, and clocked SRAM (GLB) components.
We now provide details of how each component of $C_{clk}$ is determined. We distribute the clock as 
a 4-level H-tree shown in Fig.~\ref{fig:HTree} where after every two levels the wire length reduces by a factor of 2.
The total wire capacitance of the H-tree, $C_{total-wire}$, is given by:

{\footnotesize
\begin{equation}
C_{total-wire}= \bigg[\frac{D_C}{2} + \big(\frac{D_C}{2}  \times 2\big) + \big(\frac{D_C}{4}  \times 4\big) + \big(\frac{D_C}{4}  \times 8\big)\bigg]  \times C_{w/l}
\label{eq:WireCap}
\end{equation}
}

\noindent
where $D_C$ is the chip dimension and $C_{w/l}$ is the per unit-length capacitance of the wire.
The capacitance due to clock buffer, $C_{total-buff}$, is computed as:
\begin{eqnarray}
C_{total-buff}=N_{buff}  \times C_{buff}
\label{eq:BuffCap}
\end{eqnarray}
Here, $N_{buff}$ is the number of total buffers in the H-tree and $C_{buff}$ is the input gate capacitance of a single clock buffer. In our implementation, we chose the size and number of the clock buffers to maintain a slew rate within 10\% of $T_{clk}$.
The $C_{total-PEreg}$ component of \eqref{eq:CN3} represents the capacitance due to the clocked registers in the processing elements (PEs) of the accelerator and is given by:
\begin{eqnarray}
C_{total-PEreg} = (J \times K) \times N_{FF} \times C_{FF} 
\label{eq:PERegCap}
\end{eqnarray}
Here, $(J \times K)$ is the PE array size, $N_{FF}$ is the number of 1-bit flip-flop per PE while $C_{FF}$ denotes the clocked capacitance from a single flip-flop. The clocked capacitance from the SRAM memory, $C_{SRAM}$, consists of the following components:
\begin{eqnarray}
C_{SRAM} = C_{decod} + C_{ARW-reg} + C_{BL-pre} + C_{SA-pre} 
\label{eq:CN4}
\end{eqnarray}
Here, $C_{decod}$ denotes the clocked capacitance from the decoder circuitry which comes from the synchronization of the word-line with the clock. $C_{ARW-reg}$ is the capacitance from the clocked registers (i.e., address, read, and write registers) and computed by counting the number of flip-flops in these registers. The clocked capacitance to pre-charge the bit-lines, $C_{BL-pre}$, is estimated from the number of total columns in the SRAM array. The pre-charge of each sense amplifier also needs to be synchronized with the clock, and the associated capacitance, $C_{SA-pre}$, is estimated from the number of total sense amplifiers in the SRAM array. Finally, we model
$E_{other-Cntrl}$ component of \eqref{eq:CN1} as 15\% of $E_{Layer}$ excluding $E_{DRAM}$, similar to
data from the literature.

\begin{table}[h]
\centering
\caption{Technology parameters used for CNNergy.}
\label{tbl:TechP}
\begin{tabular}{|l|c|c|c|c|c|}
\hline 
           & 16-bit MAC,         & \multicolumn{4}{c|}{Memory access, 65nm~\cite{Chen2017}} \\ \cline{3-6}
           & $\tilde{e}_{MAC}$,           & RF              & Inter-PE       & GLB           & DRAM \\
           &   45nm~\cite{Horowitz2014}          & access,       & access,        & access,      & access,\\
           &    & $\tilde{e}_{RF}$    & $\tilde{e}_{IPE}$    & $\tilde{e}_{GLB}$     & $\tilde{e}_{DRAM}$  \\ \hline
Energy     & 0.95 pJ            & 1.69 pJ     & 3.39 pJ      & 10.17 pJ    & 338.82 pJ  \\ \hline
\end{tabular}
\vspace{-0.2cm}
\end{table}

\section{Validation of CNNergy}
\label{sec:VAM}

\noindent
We validate CNNergy against limited published data for AlexNet and GoogleNet-v1:\\
(i) EyMap, the Eyeriss energy model, utilizing the mapping parameters provided
in~\cite{ChenEy2017}. This data {\em only provides parameters for the five
convolution layers of AlexNet}.\\
(ii) EyTool, Eyeriss's energy estimation tool~\cite{ETool}, excludes
$E_{Cntrl}$ and {\em supports AlexNet and GoogleNet-v1 only}.\\ 
(iii) EyChip, measured data from 65nm silicon~\cite{ChenEy2017} ({\em AlexNet
Conv layers only, excludes $E_{DRAM}$}). 

\noindent
Note that our CNNergy exceeds the capability of these: 
\begin{itemize}
\item CNNergy is suitable for customized energy access (i.e., any intermediate CNN
energy component is obtainable).
\item
CNNergy can find energy for various accelerator parameters.
\item
CNNergy can analyze a vast range of CNN topologies and general CNN accelerators, not
just Eyeriss.
\end{itemize}

To enable a direct comparison with Eyeriss, 16-bit fixed point arithmetic
precision is used to represent feature maps and filter weights. The technology
parameters are listed in Table~\ref{tbl:TechP}. The available process data is
from 45nm and 65nm nodes, and we use the factor $s = \frac{65}{45} \times \left
( \frac{V_{DD,65nm}}{V_{DD,45nm}} \right )^2$ to scale 45nm data for
direct comparison with measured 65nm silicon.

For the control energy, we model capacitances using the parameters from the NCSU 45nm process design kit (PDK)~\cite{PDK45}, the capacitive components in \eqref{eq:CN3}-\eqref{eq:CN4} are extracted to estimate $C_{clk}$.
Fig.~\ref{fig:SlewRate} shows the percent slew of the clock as we increase the load capacitance to a single clock buffer (MOSFET sizing of the buffer: length, $L = 50 nm$; width, $W_{NMOS} = 3L$ and $W_{PMOS} = 6L$). From this plot, the maximum load capacitance to each clock buffer (37 fF) is calculated to maintain a maximum of 10\% slew rate and the buffers in the H-tree are placed accordingly. 
The results are scaled to 65nm node by the scaling factor $s$. The resultant clock
power is computed by \eqref{eq:CN2}, and the {\em latency} for each layer
in \eqref{eq:CN1} is inferred as $\frac{\# of\; MAC\; per\;
layer}{Throughput}$, where the numerator is a property of the CNN topology and
the denominator is obtained from~\cite{ChenEy2017}.

Fig.~\ref{fig:AMval} compares the energy obtained from CNNergy, EyTool, and EyMap to
process an input image for AlexNet. As stated earlier, EyTool excludes
$E_{Cntrl}$; accordingly, our comparison also omits $E_{Cntrl}$. The numbers
match closely.

Fig.~\ref{fig:Wcntrl} shows the energy results for AlexNet including the
$E_{Cntrl}$ component in~\eqref{eq:Elayer} for both CNNergy and EyMap and compares the results with EyChip which represents practical energy consumption from a fabricated chip. The EyTool
data that neglects $E_{Cntrl}$ is significantly off from the more accurate data
for CNNergy, EyMap, and EyChip, particularly in the Conv layers. Due to
unavailability of reported data, the bars in Fig.~\ref{fig:Wcntrl} only show
the Conv layer energy for EyMap and EyChip.  Note that EyChip does not include
the $E_{DRAM}$ component of~\eqref{eq:Edata}.

Fig.~\ref{fig:AMGnet} compares the energy from CNNergy with the EyTool energy for
GoogleNet-v1. Note that the only available hardware data for GoogLeNet-v1 is
from EyTool, which does not report control energy: this number matches the
non-$E_{Cntrl}$ component of CNNergy closely. As expected, the energy is higher when
$E_{Cntrl}$ is included.

\section{Transmission energy and Delay Computation}
\label{sec:TrnDelay}

\subsection{Transmission Energy ($E_{Trans} $) Estimation}
\label{sec:TrnsEn}

\noindent
The transmission energy, $E_{Trans}$, is a function of the available data
bandwidth, which may vary depending on the environment that the mobile client
device is in. Similar to the prior works on offloading computation to the cloud~\cite{kang2017, Kumar2010,Nimmagadda2009} we use the following model to estimate the energy required to transmit data bits from the mobile client to the cloud.
\begin{eqnarray}
E_{Trans} = P_{Tx} \times \frac{D_{RLC}}{B_e}
\label{eq:Etran} 
\end{eqnarray}
where $P_{Tx} $ is the transmission power of the client, 
$B_e$ is the effective transmission bit rate, and $D_{RLC}$ is the number of encoded data bits to be transmitted. The time required to transmit the data bits is determined by the bit rate. Similar to~\cite{Halperin10}, the transmission power is assumed to be constant during the course of transmission after the wireless connection has been established as well as a simple fading environment is assumed.
During data transmission, typically, there is an overhead due to error correction scheme. An error correction code (ECC) effectively reduces the data bandwidth. If $k\%$ of the actual data is designated for the the ECC bits, then the effective transmission bit rate for actual data (i.e., $D_{RLC}$ in~\eqref{eq:Etran}) is given by: 
\begin{eqnarray}
B_e = \frac{B}{1 + (k/100)}
\label{eq:Beffect} 
\end{eqnarray}
where $B$ is the available transmission bit rate. For the highly sparse data at internal layers
of a CNN, run-length compression (RLC) encoding is used to reduce the
transmission overhead.  The number of transmitted RLC encoded data bits,
$D_{RLC}$, is:
\begin{eqnarray}
D_{RLC} = D_{raw} \times (1 - \mbox{{\em Sparsity}}) \times (1 + \delta)
\label{eq:Drlc}
\end{eqnarray}
Here, $D_{raw}$ is the number of output data bits at each layer including zero
elements, {\em Sparsity} is the fraction of zero elements in the respective data
volume, and $\delta$ is the average RLC encoding overhead for each bit associated with the nonzero
elements in the raw data (i.e., to encode each bit of a nonzero data element, on average, $(1 + \delta)$ bits are required). 
Using 4-bit RLC encoding (i.e., to encode information about the number of zeros between nonzero elements) for 8-bit data (for
evaluations in Section~\ref{sec:res}), and 5-bit RLC encoding for 16-bit data (during Eyeriss
validation in Section~\ref{sec:VAM}), $\delta$ is 3/5 and 1/3,
respectively (note that this overhead only applies to the few nonzeros in a
very sparse data).

\subsection{Inference Delay ($t_{delay} $) Estimation}
\label{sec:TDelay}

\noindent
Although our framework aims to optimize client energy, we also evaluate the
total time required to complete an inference ($t_{delay}$) in the client+cloud.
For a computation partitioned at the $L^{th}$ layer, the inference delay is
modeled as:
\begin{eqnarray}
\textstyle t_{delay} = \sum_{i=1}^{L} t_{client} (i) + t_{Trans} +
\sum_{i=L+1}^{|L|} t_{cloud} (i)
\label{eq:infdelay}
\end{eqnarray}
where $t_{client}(i)$ [$t_{cloud}(i)$] denote the $i^{\rm th}$ layer latency at
the client [cloud], $|L|$ is the number of layers in the CNN, and $t_{Trans}=D_{RLC}/B_e$ is the time required for
data transmission at the $L^{th}$ layer. The latency for each layer is computed as in Section~\ref{sec:VAM} where the 
{\em Throughput} comes from the client and cloud platforms.

\section{Runtime Partitioning by NeuPart}
\label{sec:RunP}

\begin{figure}[!ht]
\vspace{-0.0cm}
\centering
\includegraphics[height=4.6cm]{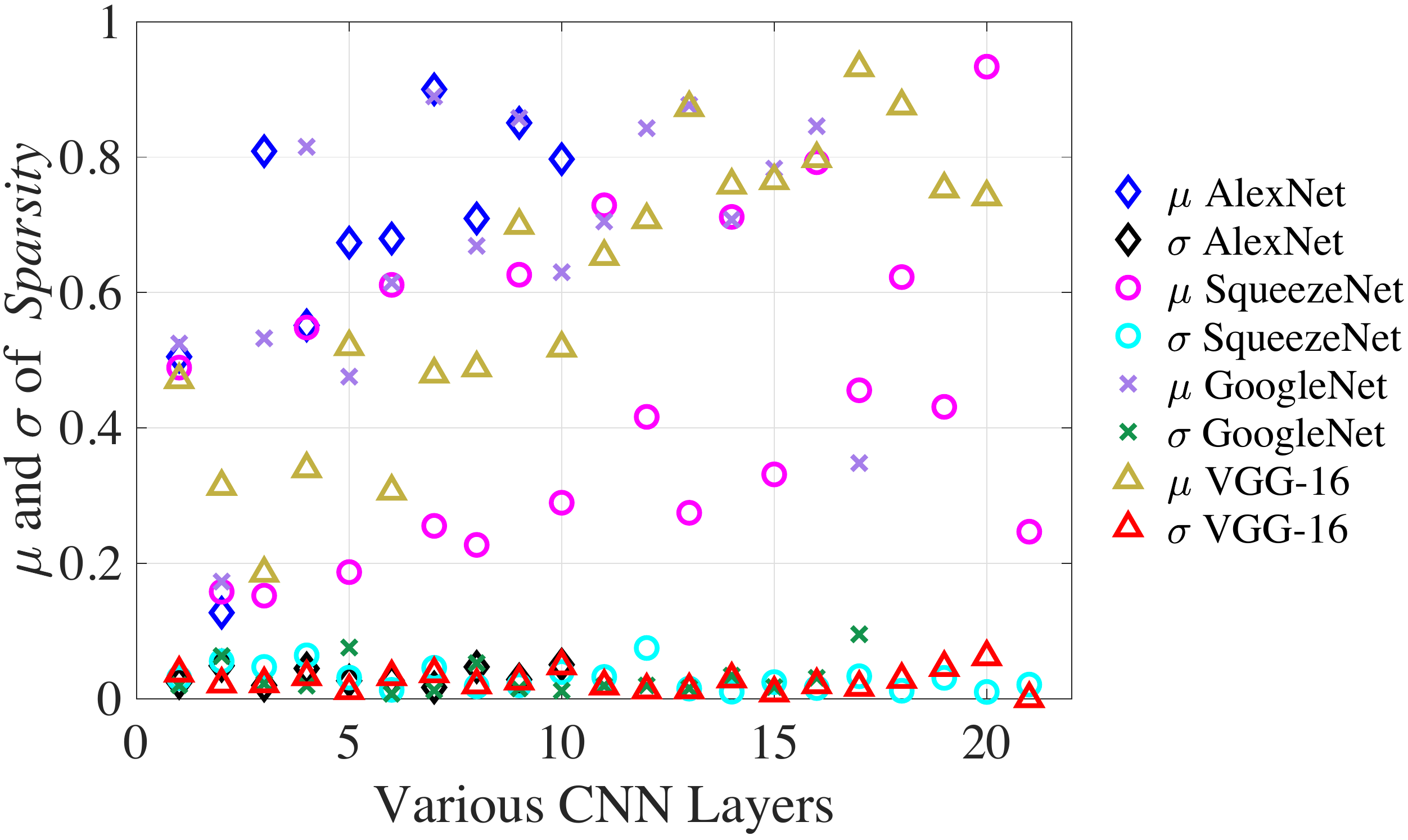}
\caption{Average ($\mu$) and standard deviation ($\sigma$) of {\em Sparsity} over
$\sim$10,000 images for AlexNet, SqueezeNet-v1.1, GoogleNet-v1, and VGG-16.
}
\label{fig:AvgStd}
\vspace{-0.0cm}
\end{figure}

\noindent
In this section, we discuss how NeuPart is used during runtime for partitioning
CNN workloads between a mobile client and the cloud.  Fig.~\ref{fig:AvgStd}
shows the average ($\mu$) and standard deviation ($\sigma$) of data sparsity at
various CNN layers over $\sim$10,000 ImageNet validation images for  AlexNet,
SqueezeNet-v1.1, GoogleNet-v1, and VGG-16. For all four networks, the standard
deviation of sparsity at all layers is an order of magnitude smaller than the
average. However, at the input layer, 
when the image is transmitted in standard JPEG compressed format, the sparsity
of the JPEG compressed image, {\em Sparsity-In}, shows significant variation
(documented in Fig.~\ref{fig:JPEGHist}), implying that the transmit energy can
vary significantly.

Therefore, a significant observation is that for all the intermediate layers,
{\em Sparsity} is primarily a feature of the network and not the input data, and can
be precomputed offline as a standard value, independent of the input image.
This implies that $D_{RLC}$, which depends on {\em Sparsity}, can be computed offline
for all the intermediate layers without incurring any optimality loss on the
partitioning decision.  Only for the input layer it is necessary to compute
$D_{RLC}$ during runtime.  The runtime optimization algorithm is therefore very
simple and summarized in Algorithm~\ref{alg:OptPart} (for notational
convenience we use superscript/subscript $L$ to indicate $L^{th}$ layer in this
algorithm).

\begin{algorithm}[h]
{\scriptsize
\textbf{INPUT:}  
$E\in \mathbb{R}^{|L|}$: Cumulative CNN energy vector;\\
$D_{RLC} \in \mathbb{R}^{|L|}$:  RLC encoded data bits for various layers;\\
$B$: Available transmission bit rate; $P_{Tx}$: Transmission power;\\
$k$: Percent overhead for ECC bits\\
\textbf{OUTPUT:} 
Optimal partition layer, $L_{opt}$
\\
\textbf{METHOD:}
\begin{algorithmic} [1]
\State 
Obtain JPEG compressed image sparsity, {\em Sparsity-In}
\label{alg:sparsity}
\State 
Update RLC data size for input layer, $D^1_{RLC}$
using {\em Sparsity-In} and \eqref{eq:Drlc}
\label{alg:InputRLC}
\State
Compute $B_e = \frac{B}{1 + (k/100)}$
\label{alg:Beffect}
\State Compute $E_{Trans}^L = P_{Tx} \times \frac{D_{RLC}^L}{B_e}$,
\hspace{0.25in} $L \in \lbrace 1, 2, \cdots |L|\rbrace$
\label{alg:Energy_transmission}
\State Compute $E_{Cost}^L = E_L + E_{Trans}^L$, \hspace{0.30in} $L \in \lbrace
1, 2, \cdots |L| \rbrace$
\label{alg:Energy_total}
\State Obtain $L_{opt}=\text{argmin}( E_{Cost} )$
\label{alg:MinFind}
\State \textbf{return $L_{opt}$}
\label{alg:2:line_final}	
\end{algorithmic}
\caption{Algorithm for Runtime Optimal Partitioning.}
\label{alg:OptPart}
}
\end{algorithm}

The cumulative CNN energy vector ($E$) up to each $L^{th}$ layer of a CNN
(i.e., $E_L = \sum_{i=1}^{L} E_{Layer}(i)$) depends on the network topology
and, therefore, precomputed offline by CNNergy. Likewise, $D_{RLC}$ for layer 2 to
$|L|$ is precharacterized using the average {\em Sparsity} value associated with each
CNN layer. During runtime, for an input image with JPEG-compressed sparsity
{\em Sparsity-In}, $D_{RLC}$ for layer 1 (i.e., input layer) is computed
(Line~\ref{alg:InputRLC}). Finally, at runtime, with a user specified
transmission bit rate $B$, percent ECC overhead $k$, and transmission power $P_{T_x}$, $E_{Cost}$ is
obtained for all the layers, and the layer that minimizes $E_{Cost}$ is
selected as the optimal partition point, $L_{opt}$
(Lines~\ref{alg:Energy_transmission}--\ref{alg:2:line_final}).

Note that both $B$ and $P_{Tx}$ are user-specified parameters in the runtime optimization algorithm.
Therefore, depending on the communication environment (i.e., signal strength, quality of the link, amount of contention from other users, variable bandwidth), a user can provide the available bit rate. Besides, depending on a specific device, the user can provide
the transmission power corresponding to that device and obtain the
partitioning decision based on the provided $B$ and $P_{Tx}$ parameters at runtime.

\noindent
{\bf Overhead of Runtime Optimization:}
The computation of Algorithm~\ref{alg:OptPart} requires only ($|L| + 1$)
multiplications, ($|L| + 2$) divisions, ($|L| + 2$) additions, and $|L|$ comparison
operations (Lines~\ref{alg:InputRLC}--\ref{alg:MinFind}), where $|L|$ is the
number of layers in the CNN topology. For standard CNNs, $|L|$ is a very small
number (e.g., for AlexNet, GoogleNet-v1, SqueezeNet-v1.1, and VGG-16, $|L|$
lies between 12 and 22). This makes NeuPart computationally very cheap to find
the optimal partition layer at runtime. Moreover, as compared to the energy required to perform the core CNN computations and data transmission, the overheard of running Algorithm~\ref{alg:OptPart} is virtually zero.

Note that the inference result returned from the cloud computation corresponds
to a trivial amount of data (i.e., only one number associated with the
identified class) which is, for example, 5 orders of magnitude lower than the
number of data bits to transmit at the P2 layer of AlexNet (already very low,
see Fig.~\ref{fig:compcomm}(b)).
Therefore, the cost of receiving the result makes no perceptible difference in
the partitioning decision.

\section{Results}
\label{sec:res}

\subsection{In Situ/Cloud Partition}
\label{sec:ClintCld}

\noindent
We now evaluate the computational partitioning scheme, using the models in
Sections~\ref{sec:AMsection} and~\ref{sec:TrnDelay}.  Similar to the state-of-the-art~\cite{Jouppi2017,
gysel2016}, we use 8-bit inference for our evaluation. The energy parameters
from Table~\ref{tbl:TechP} are quadratically scaled for multiplication and
linearly scaled for addition and memory access to obtain 8-bit energy
parameters. We compare the results of partitioning with
\begin{itemize}
\item {\bf FCC}: fully cloud-based computation
\item {\bf FISC}: fully {\em in situ} computation on the client
\end{itemize}

The energy cost $(E_{Cost})$ in \eqref{eq:Ecost} for each layer of
a CNN is analyzed under various communication environments for the mobile
cloud-connected client. 
Prior works in the literature~\cite{qian2011profiling,Altamimi2015,miettinen2010} report the measured average power of various smartphones during the uplink activity of wireless network (documented in Table~\ref{tbl:MeasPow}).
In our work, for the transmission power ($P_{Tx}$) in~\eqref{eq:Etran}, we use representative numbers from Table~\ref{tbl:MeasPow} and thus evaluate the computational partitioning scheme considering specific scenarios that correspond to specific mobile platforms.
The transmit power for an on-chip transmitter is independent of the
transmission data rate~\cite{Halperin10}, and the numbers in
Table~\ref{tbl:MeasPow} do not vary with the data rate.
We present analysis using the effective bit rate ($B_e$) as a variable parameter to evaluate the
benefit from the computation partitioning scheme as the available bandwidth
changes.  For all plots: {\em (i)}~``In'' is the input layer (i.e, the input image
data); {\em (ii)}~layers starting with ``C'', ``P'', and ``FC'' denote Conv,
Pool, and FC layer, respectively; {\em (iii)}~layers starting with ``Fs'' and ``Fe''
denote {\em squeeze} and {\em expand} layer, respectively, inside a fire module
of SqueezeNet-v1.1.

\begin{table}[htb]
\centering
\caption{Measured average power of smartphones during wireless network uplink activity.}
\label{tbl:MeasPow}
\begin{tabular}{|l|c|c|c|}
\hline
\multicolumn{1}{|c|}{\multirow{2}{*}{Wireless network}} & \multirow{2}{*}{WLAN} & \multirow{2}{*}{3G} & \multirow{2}{*}{4G LTE} \\
\multicolumn{1}{|c|}{} &  &  &  \\ \hline
Google Nexus One~\cite{qian2011profiling} & -- & 0.45 W & -- \\ \hline
LG Nexus 4~\cite{Altamimi2015} & 0. 78 W & 0.71 W & -- \\ \hline
Samsung Galaxy S3~\cite{Altamimi2015} & 0.85 W & 1.13 W & 1.13 W \\ \hline
BlackBerry Z10~\cite{Altamimi2015} & 1.14 W & 1.03 W & 1.22 W \\ \hline
Samsung Galaxy Note 3~\cite{Altamimi2015} & 1.28 W & 0.75 W & 2.3 W \\ \hline
Nokia N900~\cite{miettinen2010} & 1.1 W & 1.0 W & -- \\ \hline
\end{tabular}
\end{table}

\begin{figure}[htb]
\vspace{-0.0cm}
\centering
\subfigure[]{
\includegraphics[height=3.2cm]{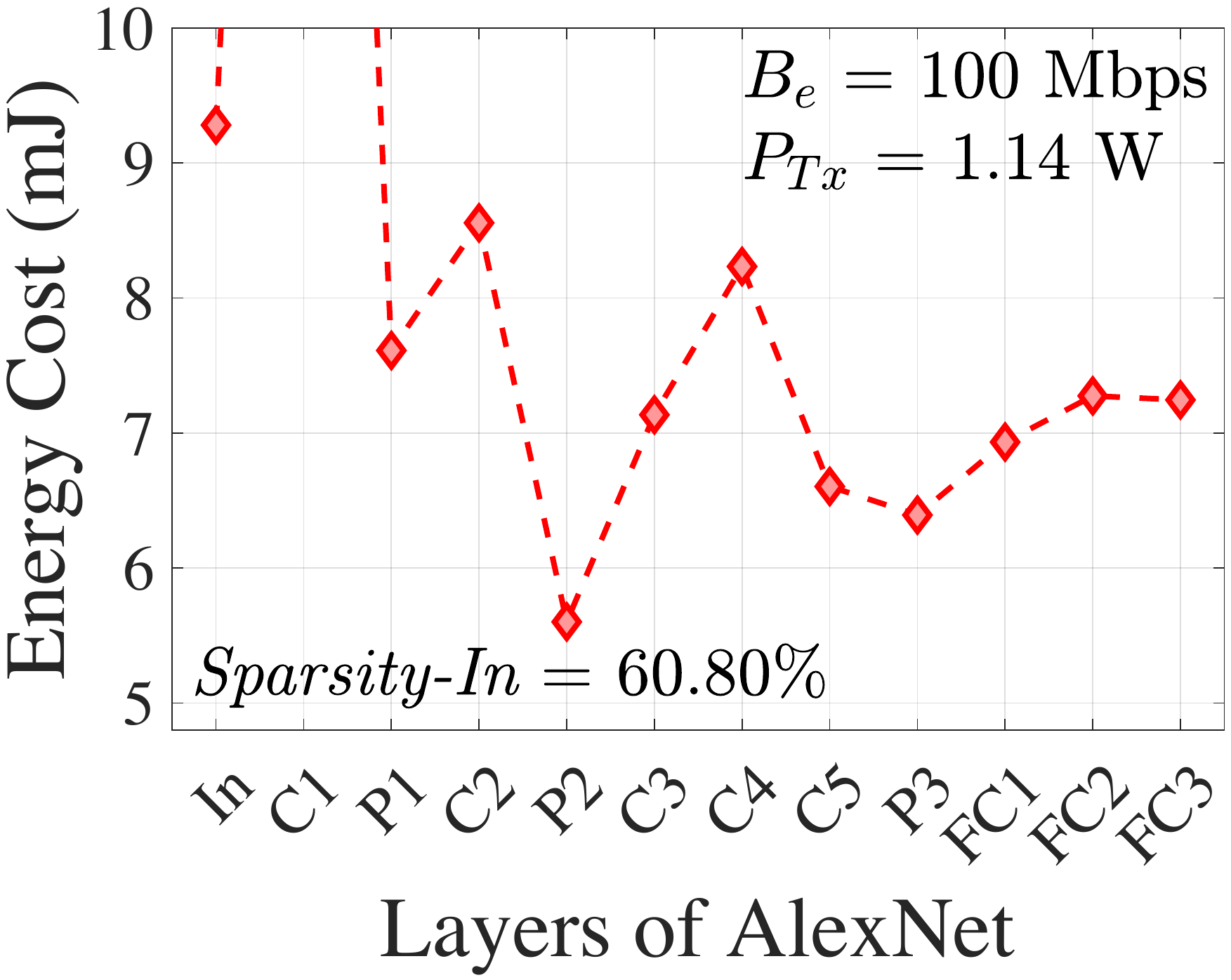}
\label{fig:Alex}
}
\subfigure[]{
\includegraphics[height=3.2cm]{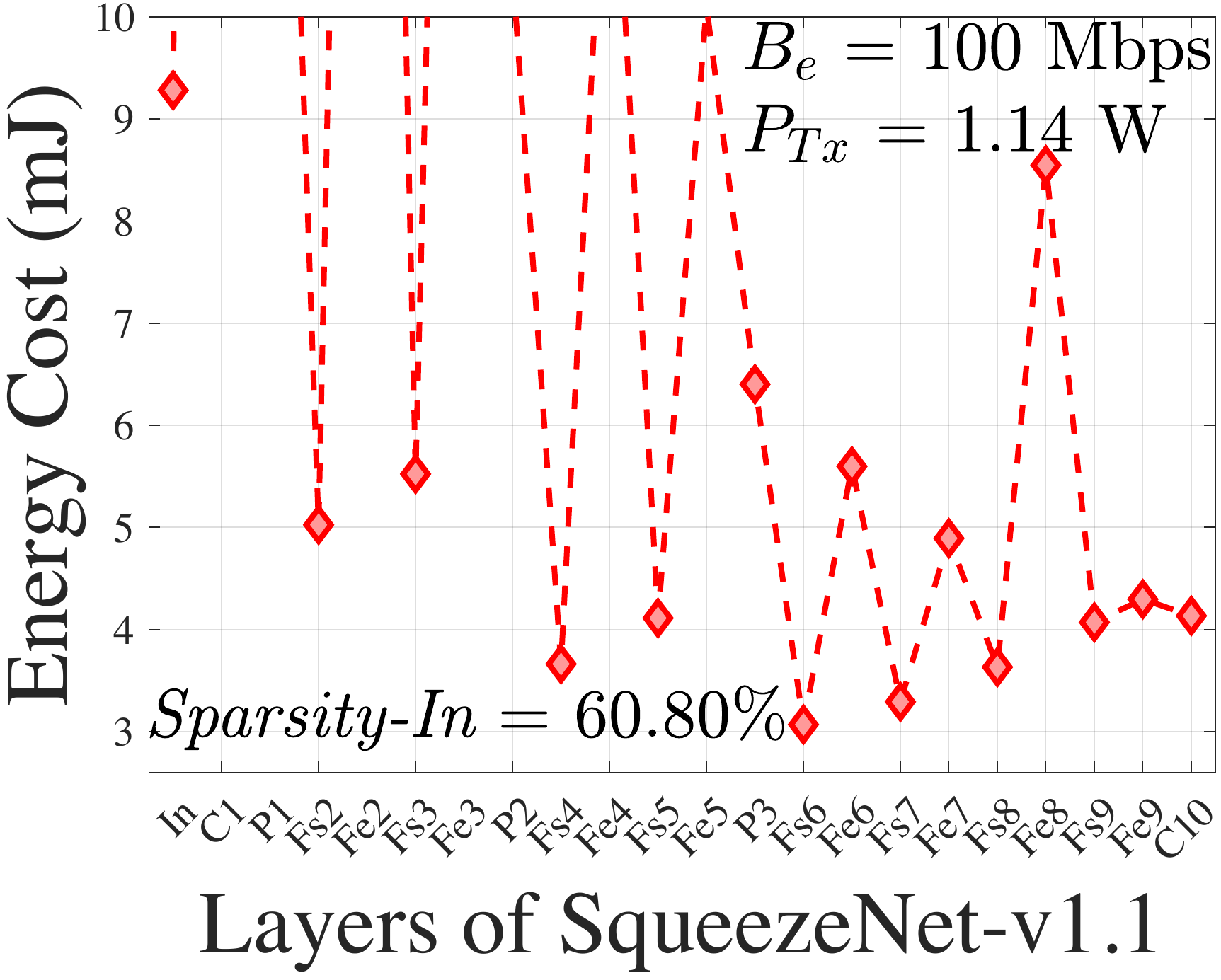}
\label{fig:SqNet}
}
\caption{Energy cost ($E_{Cost}$) at various layers for (a) AlexNet, (b) SqueezeNet-v1.1. (In both figures, points above 10 mJ are omitted for better visibility.)}
\vspace{-0.4cm}
\end{figure}

At the In layer, before transmission, the image is JPEG-compressed with a
quality level of $Q=90$ (a lower $Q$ provides greater compression but the worsened
image induces CNN accuracy degradation).  The energy overhead associated with
JPEG compression~\cite{Snigdha2016} is incorporated in $E_{Cost}$ for the In
layer but is negligible.

For an input image, Fig.~\ref{fig:Alex} shows the energy cost associated with
each layer of AlexNet at 100 Mbps effective bit rate $(B_e)$ and 1.14 W transmission power
$(P_{Tx})$ which corresponds to the BlackBerry Z10 platform. The minimum $E_{Cost}$ occurs at an intermediate layer, P2, of
AlexNet which is 39.65\% energy efficient than the In layer (FCC) and 22.7\%
energy efficient than the last layer (FISC). It is now clear that offloading
data at an intermediate layer is more energy-efficient for the client than FCC
or FISC. Using the same smartphone platform, Fig.~\ref{fig:SqNet} shows a similar result with an intermediate
optimal partitioning layer for SqueezeNet-v1.1. Here, the Fs6 layer is optimal
with an energy efficiency of 66.9\% and 25.8\% as compared to FCC and FISC,
respectively.

\begin{figure}[h!t]
\vspace{-0.4cm}
\centering
\includegraphics[height=3.8cm]{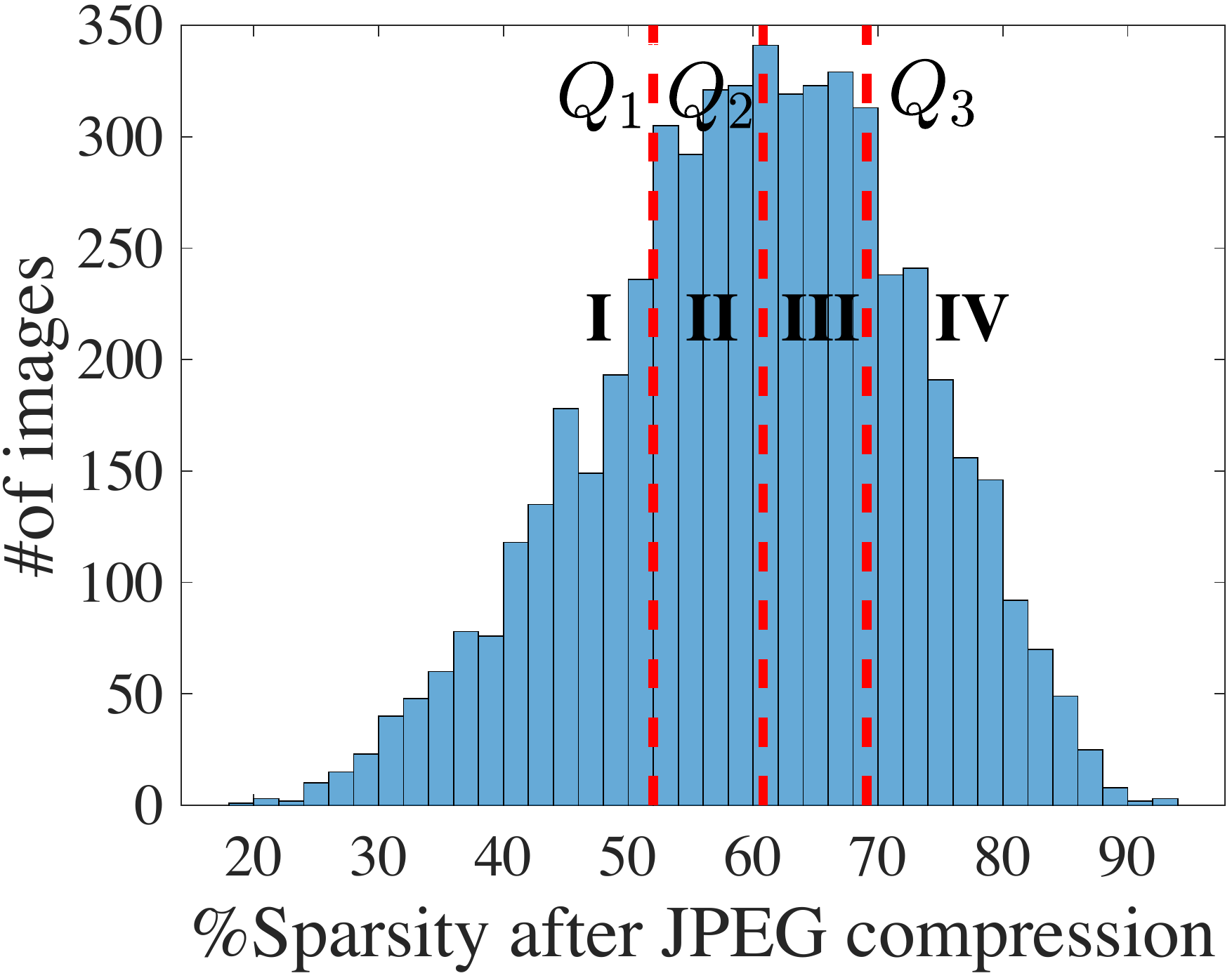}
\caption{Distribution of images with respect to {\em Sparsity-In}.}
\label{fig:JPEGHist}
\vspace{-0.4cm}
\end{figure}

The cost of FCC is image-dependent, and varies with the sparsity, {\em
Sparsity-In}, of the compressed JPEG image, which alters the transmission cost
to the cloud.  Fig.~\ref{fig:JPEGHist} shows that the $\sim$5500 test images
in the ImageNet database show large variations in {\em Sparsity-In}. We divide
this distribution into four quartiles, delimited at points $Q_1$, $Q_2$, and $Q_3$.

\begin{figure*}[b!t]
\centering
\subfigure[]{
\includegraphics[height=4.0cm]{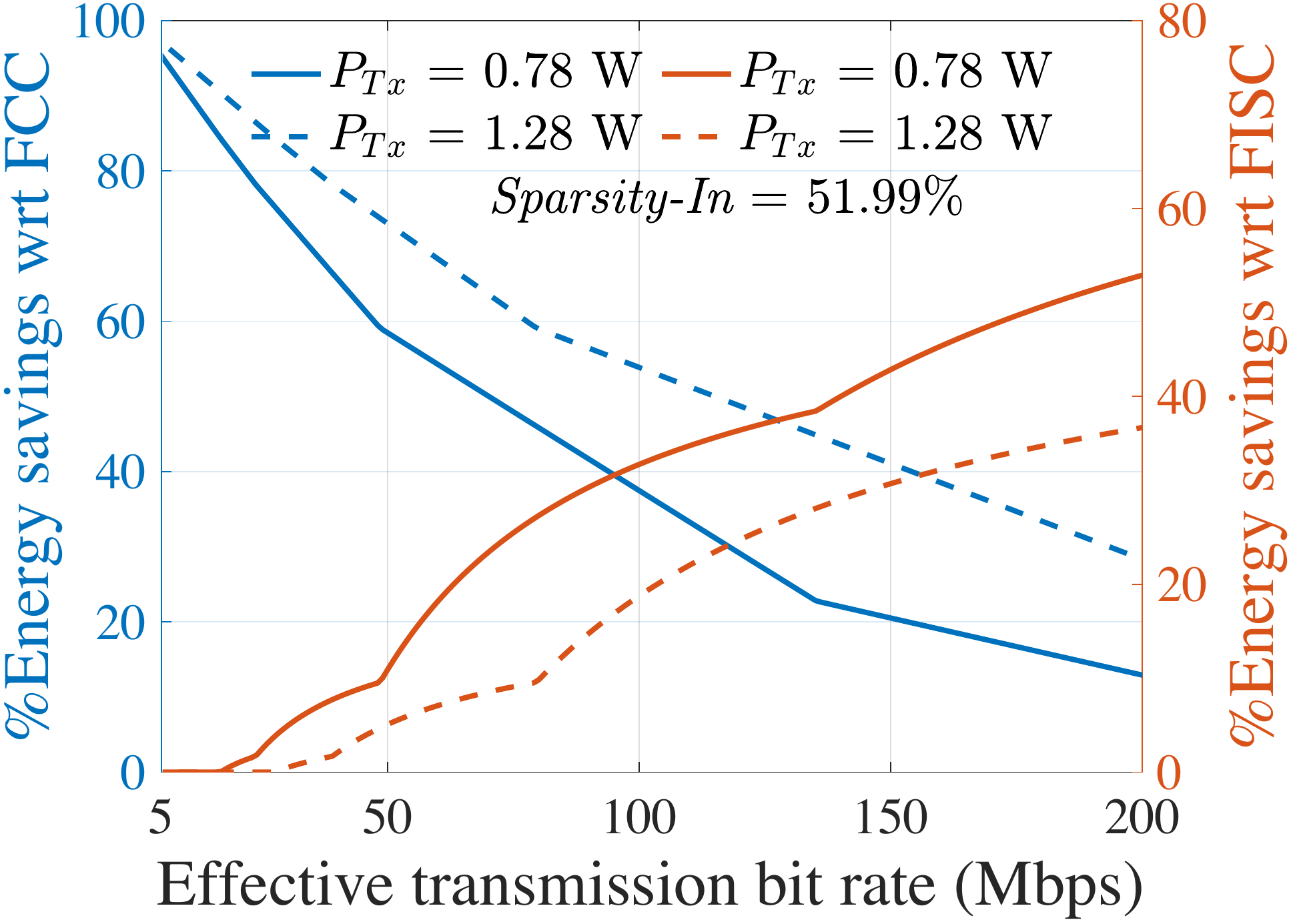}
\label{fig:Qr1}
}
\subfigure[]{
\includegraphics[height=4.0cm]{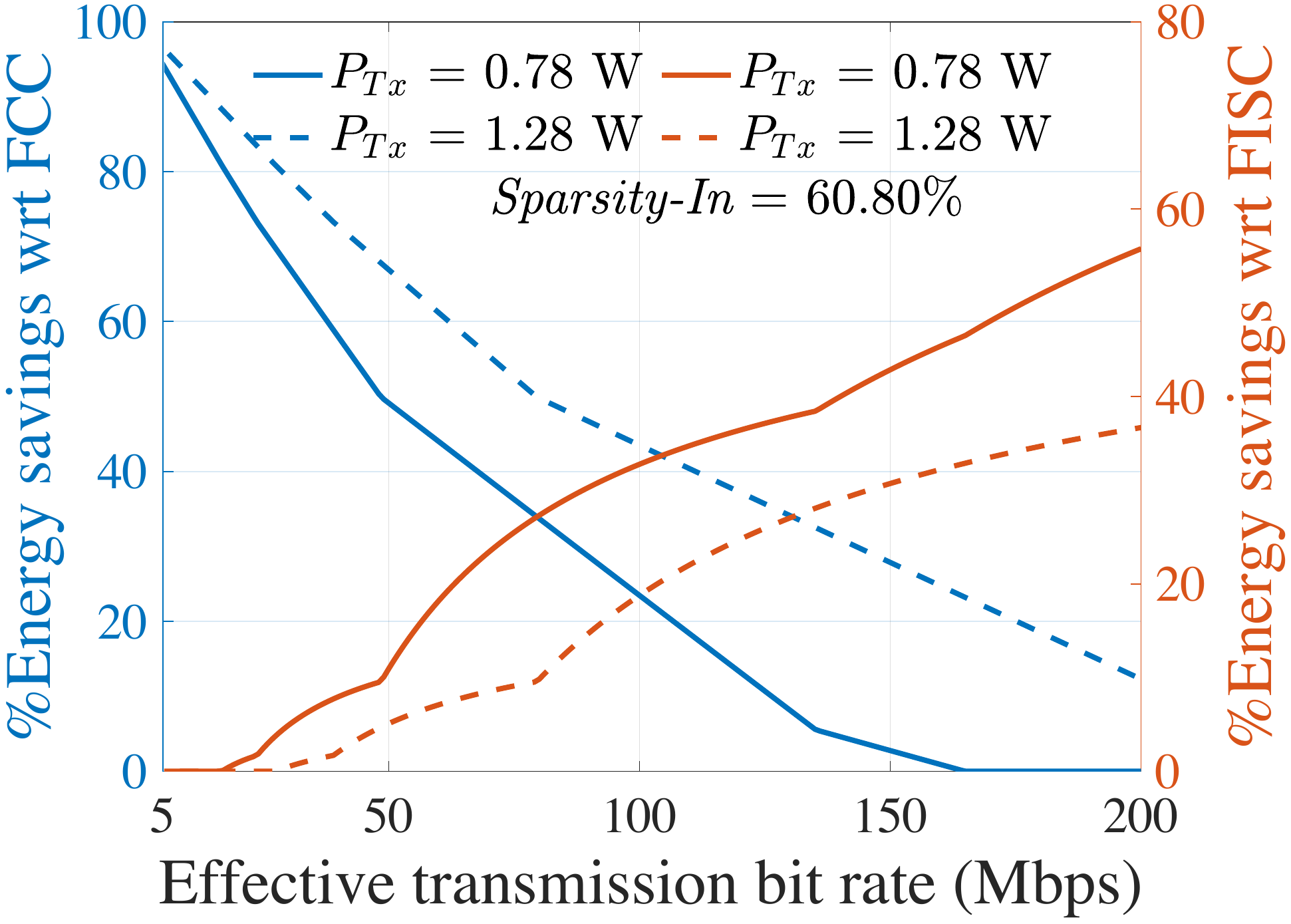}
\label{fig:Qr2}
}
\subfigure[]{
\includegraphics[height=4.0cm]{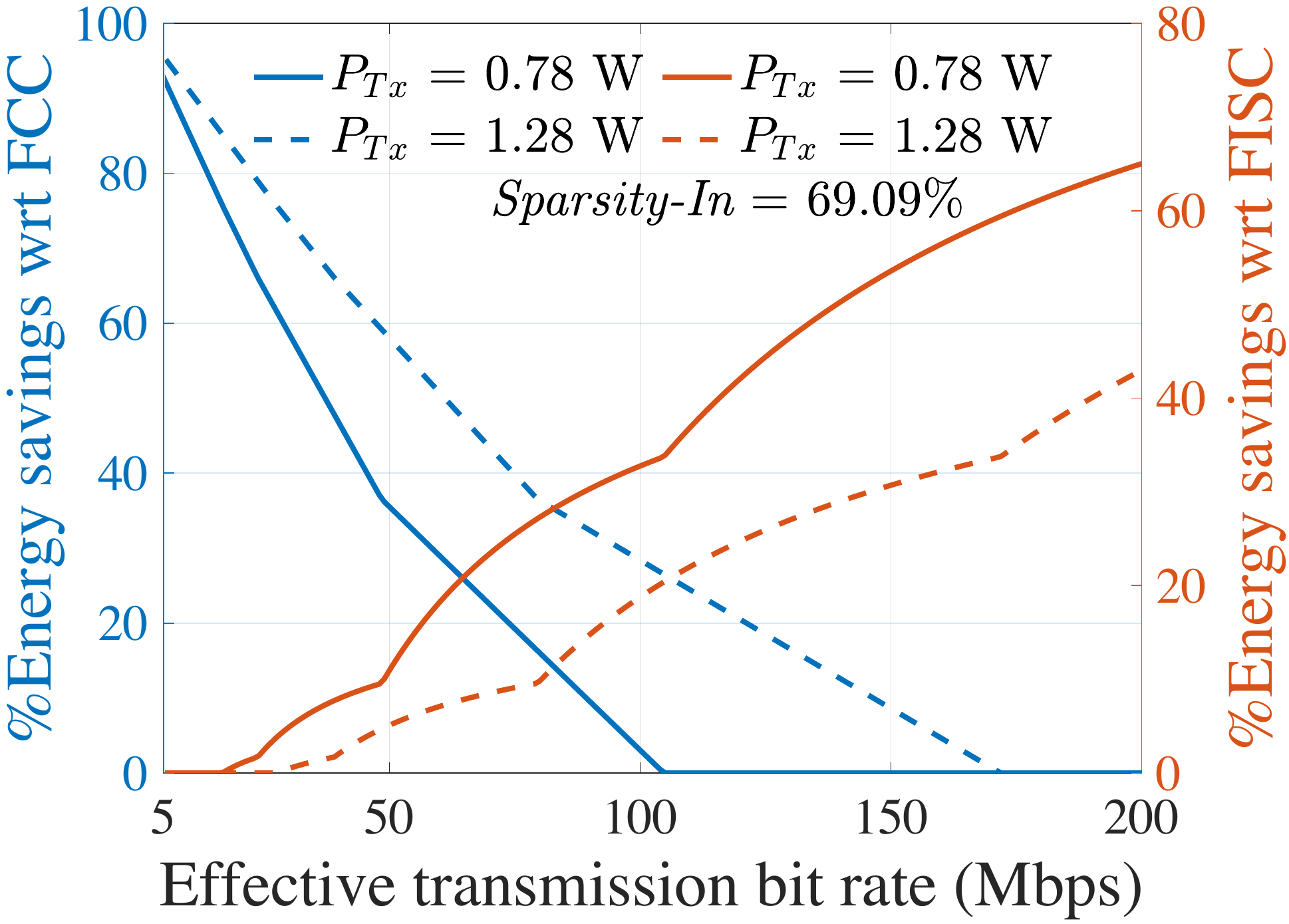}
\label{fig:Qr3}
}
\caption{Percentage energy savings of the client device under different communication environment for AlexNet at (a) {\em Sparsity-In} = 51.99\% ($Q_1$) (b) {\em Sparsity-In} = 60.80\% (Median, $Q_2$) (c) {\em Sparsity-In} = 69.09\% ($Q_3$).}
\label{fig:Quartile}
\end{figure*}

For representative images whose sparsity corresponds to $Q_1$, $Q_2$, and
$Q_3$, Fig.~\ref{fig:Quartile} shows the energy savings on the client at the
optimal partition of AlexNet, as compared to FCC (left axis) and FISC (right
axis).  For various effective bit rates ($B_e$), the plots correspond to two different $P_{Tx}$
of 0.78 W and 1.28 W, corresponding to the specifications of LG Nexus 4 and Samsung Galaxy Note 3, respectively, in Table~\ref{tbl:MeasPow}.

In Fig.~\ref{fig:Quartile}, a 0\% savings with respect to FCC [FISC] indicates
the region where the In [output] layer is optimal implying that FCC [FISC] is
the most energy-efficient choice.  Figs.~\ref{fig:Qr1} and~\ref{fig:Qr2} show
that for a wide range of communication environments, the optimal layer is an
intermediate layer and provides significant energy savings as compared to both
FCC and FISC. However, this also depends on image sparsity: a higher value of
{\em Sparsity-In} makes FCC more competitive or even optimal, especially for
images in quartile IV (Fig.~\ref{fig:Qr3}).  However, for many images in the
I-III quartiles, there is a large space where offloading neural computation at
the intermediate optimal layer is energy-optimal. 

The effect of change in $P_{Tx}$ can also be seen from the plots in Fig.~\ref{fig:Quartile}. With a higher value of $P_{Tx}$ (the dotted curves), data transmission requires more energy. Therefore, the region for which an intermediate partitioning offers energy savings (i.e., the region between energy savings of 0\% with respect to FCC and 0\% with respect to FISC) exhibits a right shift, along with a reduction in the savings with respect to FISC. However, the region also becomes wider since with a higher $P_{Tx}$, FCC becomes less competitive. With the highest $P_{Tx}$ settings from Table~\ref{tbl:MeasPow} (i.e., 2.3 W), it turns out that intermediate optimal partitioning offers limited savings with respect to FISC for a lower bit rate and from a higher bit rate (i.e., $\textgreater$ 100 Mbps) the savings starts to become considerable.
Similar trends are seen for SqueezeNet-v1.1 where the ranges of $B_e$ for which an intermediate layer is
optimal are even larger than AlexNet with higher energy savings.

The optimum partition is often, but not always, at an intermediate point for
all CNNs. For example, for GoogleNet-v1, a very deep CNN, in many cases either
FCC or FISC is energy-optimal, due to the large amount of computation as well
as the comparatively higher data dimension associated with its intermediate
layers.  However, for smaller {\em Sparsity-In} values (i.e., images which do
not compress well), the optimum can indeed occur at an intermediate layer,
implying energy savings by the client/cloud partitioning.  For VGG-16,
the optimal solution is FCC, rather than partial on-board computation
or FISC. This is not surprising: VGG-16 incurs high computation cost
and has large data volume in the deeper layers, resulting in high energy for
client side processing.

\begin{table}[htb]
\vspace{-0.0cm}
\centering
\caption{Energy savings at optimal layer for different CNN topologies ($B_e$ =
80Mbps;  $P_{Tx}$ = 0.78W for AlexNet and SqueezeNet-v1.1, 1.28W for
GoogleNet-v1).}
\label{tbl:OptStage}
\begin{tabular}{|l|r|r|r|r|l||r|}
\hline
\multicolumn{6}{|c|}{Average percent energy savings with respect to} \\ \hline 
\multicolumn{5}{|c|}{FCC}                                                                                                                                                                                                                                                & FISC \\ \cline{1-5} 
\multirow{2}{*}{CNN} & \multicolumn{4}{c|}{Quartile} & 
\\ \cline{2-5}
           &  I       & II      & III     & IV       &  
\\ \hline
AlexNet    & \cellcolor{green!20} 52.4\%   & \cellcolor{green!20}  40.1\%  & \cellcolor{green!20} 25.7\%  & \cellcolor{green!20} 4.1\%    & \cellcolor{green!20} 27.3\% 
\\ \hline
SqueezeNet & \cellcolor{green!20} 73.4\%   & \cellcolor{green!20} 66.5\%  & \cellcolor{green!20} 58.4\%  & \cellcolor{green!20} 38.4\%   & \cellcolor{green!20} 28.8\% 
\\ \hline
GoogleNet  & \cellcolor{green!20} 21.4\%   & \cellcolor{green!20} 3.5\%   & 0.0\%   & 0.0\%    & \cellcolor{green!20} 10.6\%
\\ \hline
\end{tabular}
\end{table}

\begin{figure*}[b!t]
\centering
\subfigure[]{
\includegraphics[height=4.0cm]{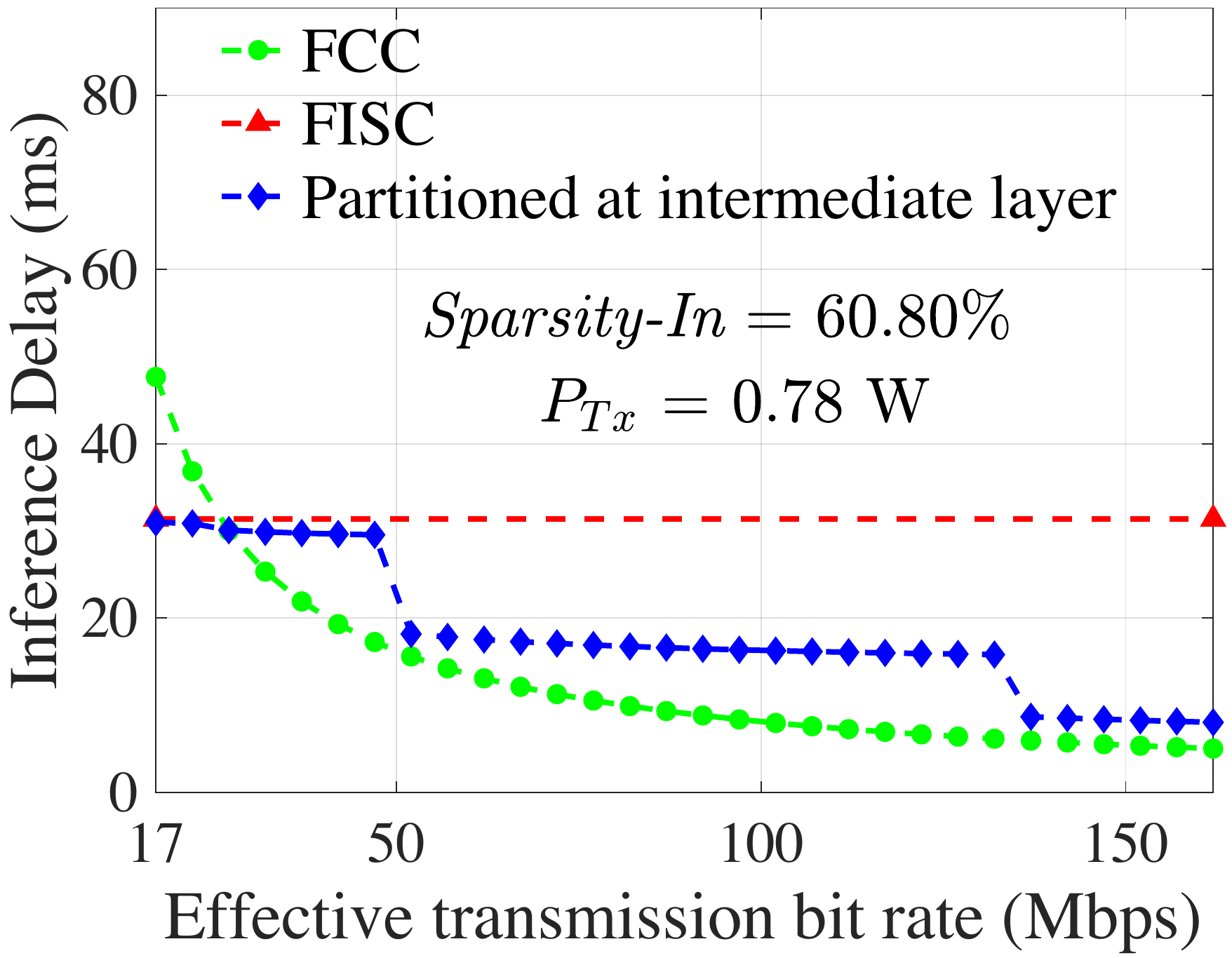}
\label{fig:IDelay}
}
\subfigure[]{
\includegraphics[height=4.0cm]{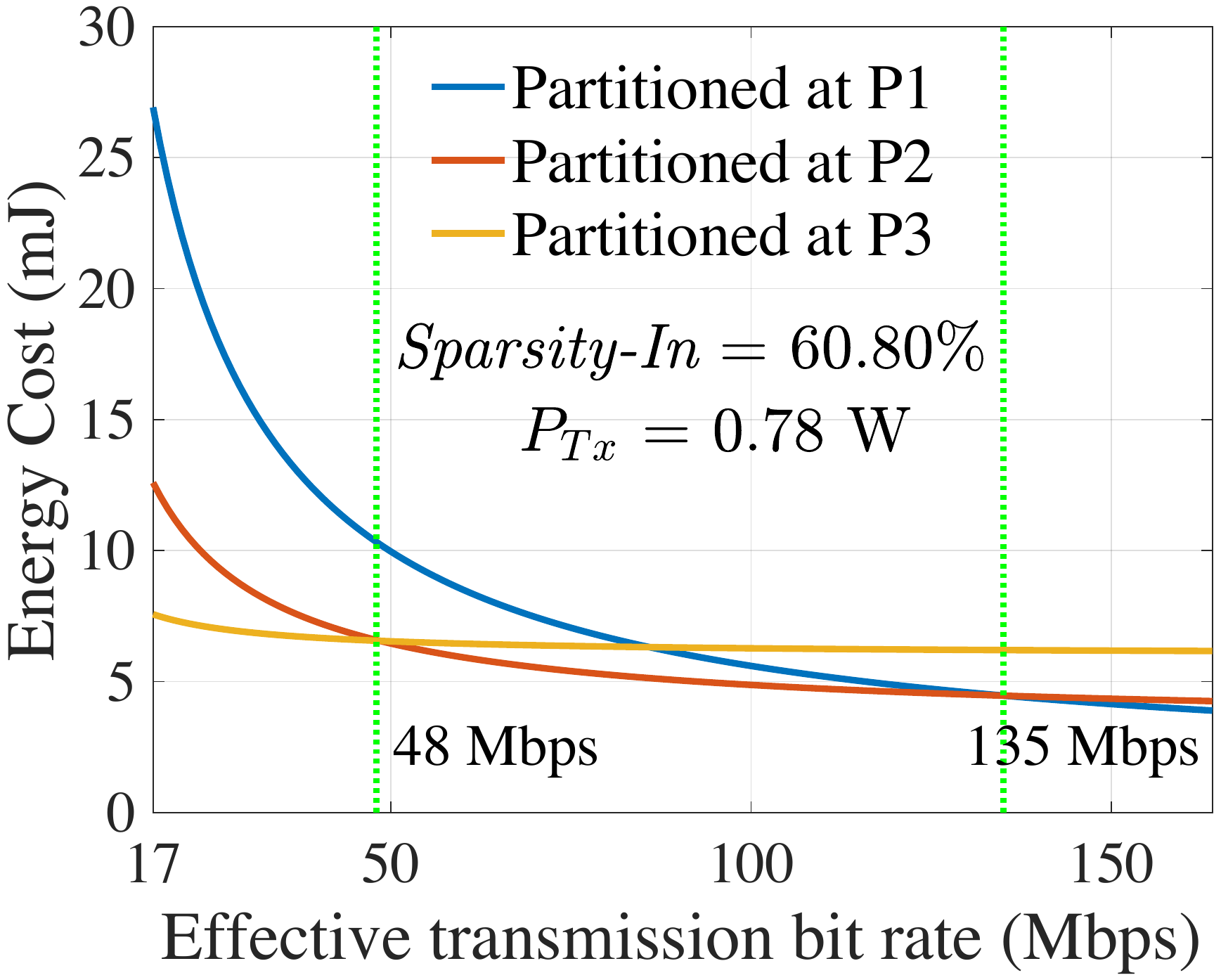}
\label{fig:BVariation}
}
\subfigure[]{
\includegraphics[height=4.0cm]{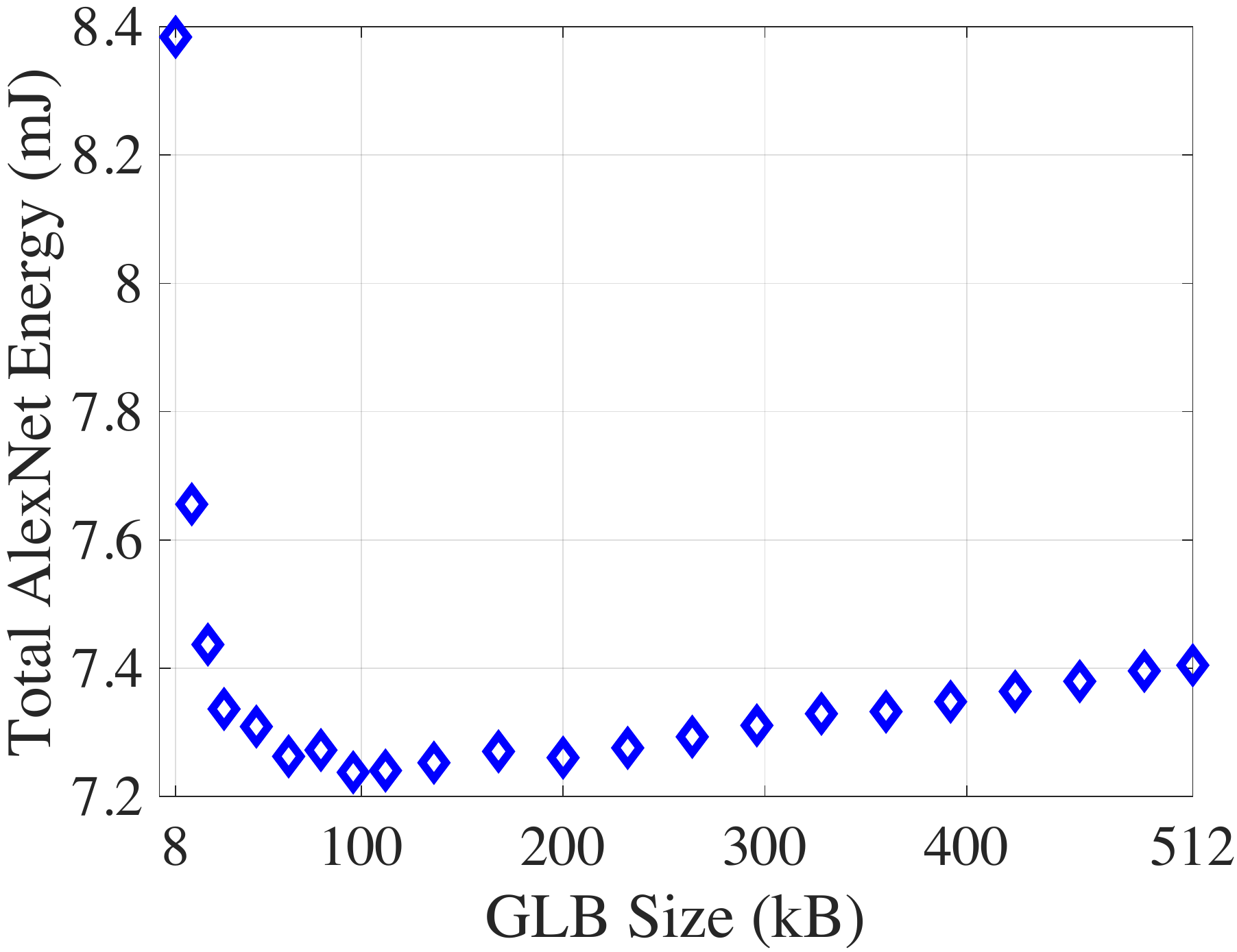}
\label{fig:GLBSize}
}
\caption{Evaluations on AlexNet: (a) Inference delay ($t_{delay}$) with respect
to FCC and FISC while partitioned at energy-optimal intermediate layers.
(b) Energy cost with variation in effective bit rate ($B_e$) when partitioned at P1,
P2, and P3 layers. 
(c) Total AlexNet energy vs. GLB size.}
\label{fig:AlexNetEval}
\end{figure*}

Under a fixed transmission power (corresponding to the platforms of LG Nexus 4 and Samsung Galaxy Note 3) and bit rate, Table~\ref{tbl:OptStage} reports the average energy savings at the optimal layer as compared to FCC and FISC for
all the images lying in Quartiles I--IV, specified in Fig.~\ref{fig:JPEGHist}.
Note that the savings with respect to FISC do not depend on {\em Sparsity-In}.
The shaded regions in Table~\ref{tbl:OptStage} indicate the regions where
energy saving is obtained by the client/cloud partitioning. For AlexNet, the
optimum occurs at an intermediate layer mostly for the images in Quartiles
I--III while providing up to 52.4\% average energy savings. For
SqueezeNet-v1.1, in all four quartiles, most images show an optimum at an
intermediate layer and provide up to 73.4\% average energy savings on the client.

\noindent
{\bf \textit{Evaluation of Inference Delay:}}
To evaluate the inference delay ($t_{delay}$), we use
GoogleTPU~\cite{Jouppi2017}, a widely deployed DNN accelerator in datacenters,
as the cloud platform, with $t_{cloud}$ in \eqref{eq:infdelay} use {\em
Throughput} = 92 TeraOps/s. At the median {\em Sparsity-In} value ($Q_2$),
Fig.~\ref{fig:IDelay} compares the $t_{delay}$ of energy-optimal partitioning
of AlexNet with FCC and FISC for various effective bit rate. The delay of FISC does not
depend on communication environment and exhibits a constant value whereas the
delay of FCC reduces with higher bit rate. The range of $B_e$ for which an
intermediate layer becomes energy-optimal is extracted using $Q_2$
(Fig.~\ref{fig:Qr2}).  The blue curve in Fig.~\ref{fig:IDelay} shows the
inference delay when partitioned at those energy-optimal intermediate layers. 
At 49 Mbps and 136 Mbps the curve shows a step reduction in delay since at these points the optimal layer shifts from P3 to P2 and from P2 to P1, respectively. It is evident from the figure that in terms of inference delay, energy-optimal intermediate layers
are either better than FCC (lower bit-rate) or closely follow FCC (higher
bit-rate) and most cases are better than FISC.

\noindent
{\bf \textit{Impact of Variations in $B$}}: We have analyzed the impact of
changes in the available bandwidth $B$ (e.g., due to network crowding) on the
optimal partition point.  For an image with {\em Sparsity-In} of $Q_2$ and 0.78
W $P_{Tx}$, Fig.~\ref{fig:BVariation} shows the energy cost of AlexNet when
partitioned at P1, P2, and P3 layers (the candidate layers for an intermediate
optimal partitioning). It shows that the energy valley is very flat with
respect to bit rate when the minimum $E_{Cost}$ shifts from P3 to P2 and from P2 to
P1 layer (the green vertical lines).  Therefore, changes in bit rate negligibly
change energy gains from computational partitioning.  For example, in
Fig.~\ref{fig:BVariation}, layer P3 is optimal for $B_e = 17-48$ Mbps, P2 is
optimal for $B_e=49-135$ Mbps, and P1 is optimal for $B_e=136-164$ Mbps. However, if
$B_e$ changes from 130 to 145 Mbps, even though the optimal layer changes from P2
to P1, the energy for partitioning at P2 instead of P1 is virtually the same.

\subsection{Design Space Exploration Using CNNergy}
\label{sec:DesignSp}
\noindent
We show how our analytical CNN energy model (CNNergy) in
Section~\ref{sec:AMsection} can be used to perform design space exploration
of the CNN hardware accelerator. For the 8-bit inference on an AlexNet
workload, Fig~\ref{fig:GLBSize} shows the total energy as a function of the
global SRAM buffer (GLB) size.  The GLB energy vs. size trend was extracted
using CACTI~\cite{CACTI}. 

When the GLB size is low, data reuse becomes difficult since the GLB can only
hold a small chunk of ifmap and psum at a time. This leads to much higher total
energy. As the GLB size is increased, data reuse improved until it saturates.
Beyond a point, the energy increases due to higher GLB access cost.
The minimum energy occurs at a size of 88kB.  However, a good engineering
solution is 32kB because it saves 63.6\% memory cost over the optimum, 
with only a 2\% optimality loss. Our CNNergy supports similar design space
exploration for other accelerator parameters as well.

\section{Conclusion}
\label{sec:Conclu}

\noindent
In order to best utilize the battery-limited resources of a cloud-connected
mobile client, this paper presents an energy-optimal DL scheme that uses
partial {\em in situ} execution on the mobile platform, followed by data
transmission to the cloud. An accurate analytical model for CNN energy (CNNergy) has been
developed by incorporating implementation-specific details of a DL accelerator
architecture. To estimate the energy for any CNN topology on this accelerator,
an automated computation scheduling scheme is developed, and it is shown to
match the performance of the layer-wise ad hoc scheduling approach of prior
work~\cite{ChenEy2017}. The analytical framework is used to predict the energy-optimal partition point for mobile client at runtime, while executing CNN workloads, with an efficient algorithm. The {\em in situ}/cloud partitioning
scheme is also evaluated under various communication scenarios. The evaluation
results demonstrate that there exists a wide communication space for AlexNet and
SqueezeNet where energy-optimal partitioning can provide remarkable energy
savings on the client.


\ifCLASSOPTIONcaptionsoff
  \newpage
\fi



\bibliographystyle{IEEEtran}
\bibliography{bib/main}



\end{document}